\documentclass[english,aps,twocolumn,groupaddress,showpacs,pre]{revtex4}
\usepackage[T1]{fontenc}
\usepackage[latin9]{inputenc}
\usepackage{textcomp}
\usepackage{amstext}
\usepackage{graphicx}
\usepackage{amssymb}
\usepackage{esint}

\makeatletter
\@ifundefined{textcolor}{}
{%
 \definecolor{BLACK}{gray}{0}
 \definecolor{WHITE}{gray}{1}
 \definecolor{RED}{rgb}{1,0,0}
 \definecolor{GREEN}{rgb}{0,1,0}
 \definecolor{BLUE}{rgb}{0,0,1}
 \definecolor{CYAN}{cmyk}{1,0,0,0}
 \definecolor{MAGENTA}{cmyk}{0,1,0,0}
 \definecolor{YELLOW}{cmyk}{0,0,1,0}
 }

\usepackage{times}
\renewcommand{\citet}{\cite}
\usepackage{mathptmx}
\usepackage{graphicx}
\usepackage{bm}

\makeatother

\usepackage{babel}

\begin{document}

\title{Solitons in graphene}

\author{Jigger Cheh$^{1}$}

\email{jk_jigger@xmu.edu.cn}

\author{Hong Zhao$^{1,2}$}

\email{zhaoh@xmu.edu.cn}

\affiliation{$^{1}$Department of Physics, Institute of Theoretical Physics and
Astrophysics, Xiamen University, Xiamen 361005, China}

\affiliation{$^{2}$State Key Laboratory for Nonlinear Mechanics, Institute of
Mechanics, Chinese Academy of Sciences, Beijing 100080, China}

\begin{abstract}
In this paper we demonstrate the direct evidence of solitons in graphene
by means of molecular dynamics simulations and mathematical analysis.
It shows various solitons emerge in the graphene flakes with two different
chiralities by cooling procedures. They are in-plane longitudinal
and transverse solitons. Their propagations and collisions are studied
in details. A soliton solution is derived by making several valid
simplifications. We hope it shed light on understanding the unusual
thermal properties of graphene.
\end{abstract}
\maketitle
Solitons are localized particle-like wavepackets which preserve their
identities such as shape and amplitude during propagation and after
collision between them due to the compensation of nonlinearity for
dispersion. Nowadays, solitons are under intense investigation in
many systems including Bose-Einstein condensates, nonlinear optics,
plasmas and anharmonic lattices\citet{1.Bose Einstein,2.nonoptics,3.plasma,4.Zhao,5.Zhao.Only},
etc.

The role solitons played in thermal properties in low dimensional
lattices such as thermal rectification and divergence of thermal conductivity
has been speculated for a long time. Thermal conductivity would be
direction dependent if solitons are involved. It was first theoretically
introduced in 1992\citet{6.1992soliton} and experimentally observed
in asymmetrically mass-loaded carbon nanotubes\citet{7.rectifier CNT}
where the rectification coefficient is estimated by considering KdV
solitons. Similar thermal rectification has been numerically studied
in asymmetric shaped graphene\citet{8.rectification nano and LiBaowen}
recently. Anomalous thermal conduction that thermal conductivity depends
on the size of the system has been investigated for years\citet{4.Zhao,5.Zhao.Only,9.anomalous thermal condution(Xiong),10.review low,11.review Dhar}.
Different theoretical approaches are applied to understand its mechanism\citet{12.momentum 2000,13.Energy Diffusion Li,14.EPL ka}.
Since solitons appear in those systems with anomalous transport properties,
it is extremely important to assess whether such properties are related
to solitons. Solitons are invoked to explain thermal divergence dates
back to Toda in 1972\citet{10.review low,15.Toda}. Later it shows
solitons might be responsible energy carriers for thermal divergence\citet{16.soliton_PRE_PRL}
and the divergent exponent is dependent upon the scattering rates
of solitons and soliton-phonon coupled energy diffusion process\citet{4.Zhao,5.Zhao.Only}.
Since anomalous thermal conduction was observed in both carbon nanotubes
and graphene flakes\citet{17.CNT break down,18.size graphene}, thus
it is very important to investigate whether solitons exist in graphene.%
\begin{figure}
\includegraphics[scale=0.3]{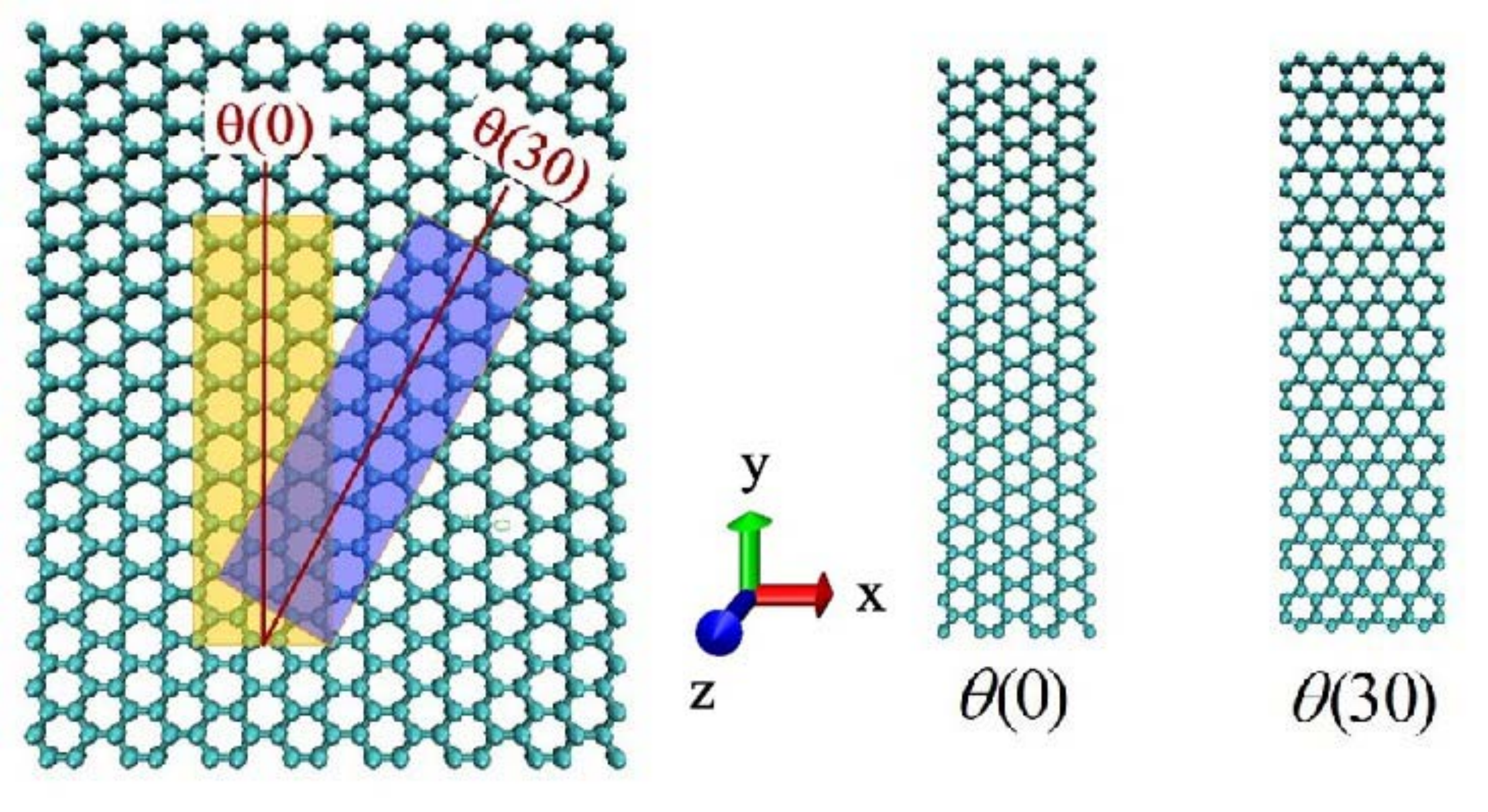}\caption{(Color online) Chiralities of the rectangle graphene flakes. Graphene
$\theta(0)$ denotes the graphene flake with armchair edge along the
x-axis and zigzag edge along the y-axis (orange flake). Since rotate
Graphene $\theta(0)$ clockwise by 30\textdegree{} we get a new rectangle
graphene flake with armchair edge and zigzag edge swapped (blue flake)
thus we denote it as Graphene $\theta(30)$. }

\end{figure}

Though the thermal properties indicate solitons might exist in graphene
however there is no direct evidence so far despite some theoretical
work suggests the possibility of supersonic KdV solitons in carbon
nanotubes and graphite\citet{19.nanotube KDV,20.supersonic KDV graphite}.
Recently some paper surmises the existence of possible solitons in
graphene under specific conditions or configurations. Five types of
solitary waves are numerically found possible along the hydrogen-terminated
edge of graphene which corresponds to the molecule group CH\citet{21.surface soliton}.
When strain is applied spatially localized electric state might appear
near the edge of graphene\citet{22. strained soliton}. It is connected
to a soliton solution in chiral gauge theory which describes the special
state of electrons. It does not describe the lattice vibration modes.
Thus in general case whether solitons would exist in graphene is still
unanswered.

In this paper we demonstrate the emergence of solitons in graphene
in both molecular dynamics simulations and mathematical analysis.
It shows various solitons emerge in graphene flakes with two different
chiralities by cooling procedures. They are in-plane longitudinal
and transverse solitons. They preserve their identities such as shape
and amplitude during propagations and after collisions. Their velocities
are smaller than the relative sound speeds. Phase shifts are brought
by an averaged acceleration effect in collision between two longitudinal
or two transverse solitons. A longitudinal and a transverse soliton
would simply pass through each other without any interactions. We
derive a NLS (Nonlinear Schr{\"o}dinger) equation to obtain the analytical
soliton solution with several simplifications. The validity of the
analytical results is discussed by comparing with the simulation results.%
\begin{figure}
\includegraphics[scale=0.2]{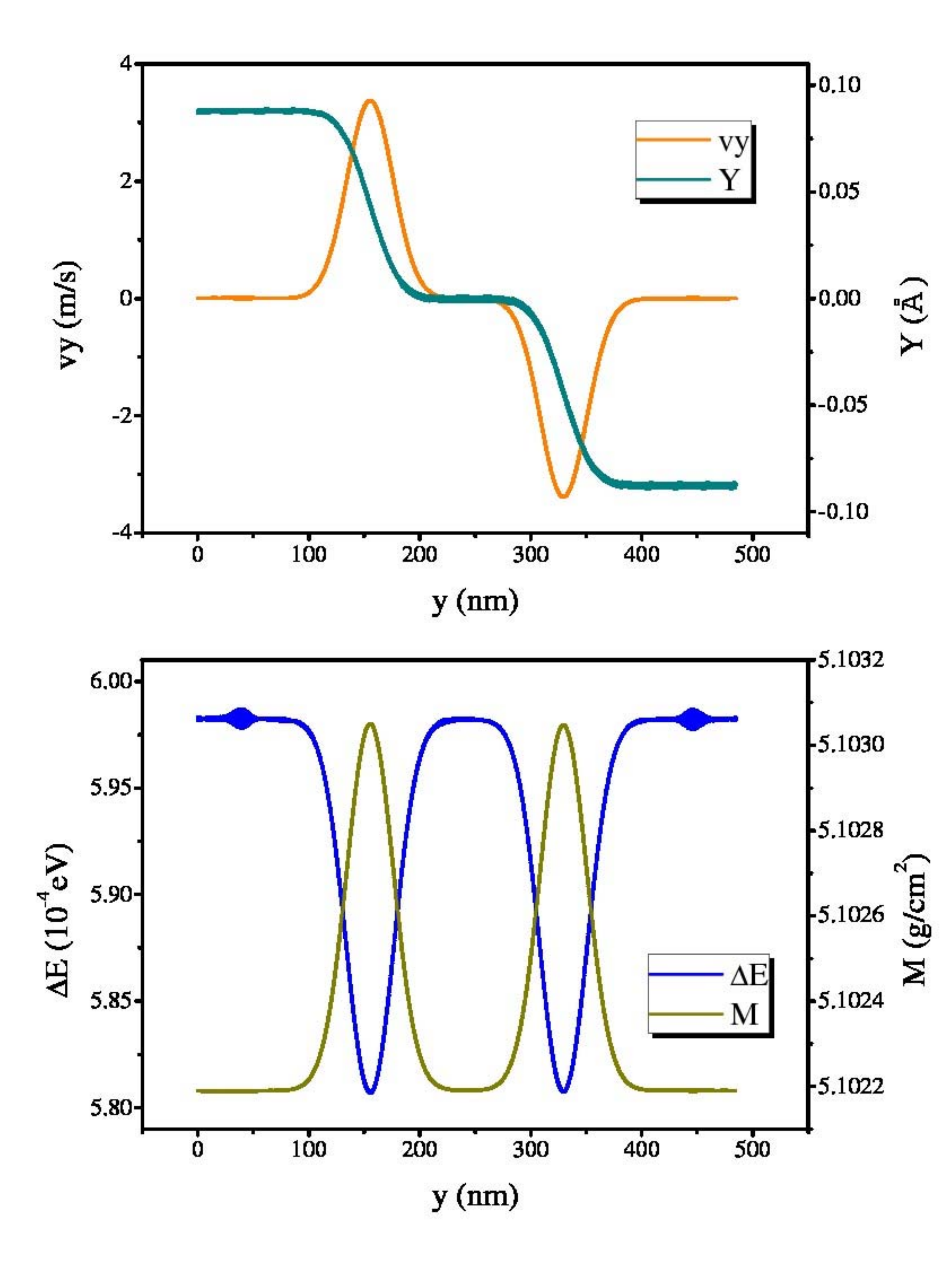}\caption{(Color online) Typical solitons in Graphene $\theta(0)$ are investigated.
Above it shows two longitudinal solitons and their velocity and displacement
distributions. Down it shows their energy and density distributions.
A pair of small breather-like solitons can also be identified.}

\end{figure}

We carry out the simulations in two rectangle graphene flakes with
different chiralities. As shown in Fig. 1, Graphene flake with armchair
edge along the x-axis and zigzag edge along the y-axis is denoted
as $\theta(0)$ and Graphene flake with zigzag edge along the x-axis
and armchair edge along the y-axis is denoted as $\theta(30)$. Graphene
$\theta(0)$ contains 95952 atoms and Graphene $\theta(30)$ contains
97020 atoms. They are 485 nm long along the y-axis and 4.7 nm wide
along the x-axis. Periodic boundary condition is used in the y-axis
which allows solitons to propagate arbitrary long distance without
boundary scattering. Periodic or free boundary condition is used in
the x-axis. Free boundary condition is used in the z-axis when concerned. 

We use the second-generation reactive empirical bond-order (REBO)
potential\citet{23.REBO2nd} as implemented in the LAMMPS\citet{24.LAMMPS}
code to simulate the anharmonic coupling between carbon atoms. We
take 0.25 fs as the minimum timestep. The second-generation reactive
empirical bond-order (REBO) potential\citet{23.REBO2nd} originates
from the Brenner\textquoteright{}s potential\citet{25.Brenner OR}
by including modified fitting data and empirical parameters. The general
form of both potentials can be written as: 

$\,$

$E_{ij}=V_{ij}^{R}(r_{ij})+b_{ij}V_{ij}^{A}(r_{ij})$ $\qquad\qquad\qquad\qquad\qquad\qquad$(1)

$\,$

Here $V_{ij}^{R}$ and $V_{ij}^{A}$ are repulsive and attractive
pairwise potential of carbon atoms dependent upon their distance.
$b_{ij}$ is a many-body bond-order term dependent upon their bond
angles and neighbor atoms. They are widely used to study diamond,
fullerene, carbon nanotube and graphene, etc. Both potentials are
capable of describing the short-ranged covalent-bonding C-C interactions
of the carbon atoms in the same way with minor differences. It should
be emphasized that, despite the effective treatment of hybridization
and zero-point energy, both potentials include no explicit quantum
effects. All conjugation states are derived entirely from the system
geometry and treat the electronic degrees of freedom in a purely empirical
manner. Thus all the simulation results are obtained and discussed
in the framework of classic mechanics.

We also test our simulations by using Tersoff potential\citet{26.Tersoff}
whichi is also a widely used empirical potential for the carbon atoms.
Similar results would be obtained. It states the main results are
general for the potentials describing the nonlinear interactions of
carbon atoms.

To identify the solitons in graphene a new strategy is used in our
simulations. Later we shall explain why such strategy works. We first
thermalize the graphene flakes at a given temperature (e. g., 1-30
K) for 100 ps with Nose-Hoover thermal bath and relax for 20 ps to
be equilibrated. Then we set the velocities of all atoms to zero every
constant step (e. g., 100-500 fs) for hundreds of times. After those
procedures various solitons emerge from graphene with different amplitudes
dependent upon the initial temperature and cooling time. Using different
initial conditions and time steps, the emerging solitons might be
different. 

We focus upon several different distributions of atoms to illustrate
the identities of the emerging solitons. Velocity distributions of
atoms are investigated which include longitudinal velocity (vy) distributions
and transverse velocity (vx) distributions. They also stand for the
momentum distributions when the mass of the carbon atoms is multiplied.
Displacements around the equilibrium positions of the atoms which
include longitudinal displacement (Y) and transverse displacement
(X) are investigated. Energy distributions are investigated by considering
the differences between the ground state energy ($\Delta E=E-E_{0}$).
Density distributions are investigated by considering the reduced
density M rather than density $\rho$ for simplicity. The relation
between them is:

$\,$

$\rho=1.99\times\frac{4\sqrt{3}}{9}M=1.99\times\frac{4\sqrt{3}}{9}\times\frac{1}{<r>^{2}}$
$\,$$\qquad\qquad\qquad$(2)

$\,$

Here $<r>$ is the average bond distance of an atom between its three
neighbors.

The typical solitons in Graphene $\theta(0)$ are shown in Fig. 2-4.
A pair of longitudinal solitons and a pair of breather-like solitons
are shown in Fig. 2 and Fig. 3. A pair of longitudinal solitons and
another pair of small longitudinal solitons are shown in Fig. 4. 

The velocity and displacement distributions of the two main longitudinal
solitons are shown in Fig. 2. Their energy and density distributions
are also shown in Fig. 2. They lack energy and aggregate density towards
the fluctuation background. A pair of small breather-like solitons
is also identified as shown in Fig. 3. Using different initial conditions
another pair of small longitudinal solitons can be identified. It
is shown in Fig. 4 they aggregate energy and lack density towards
the fluctuation background.%
\begin{figure}
\includegraphics[scale=0.2]{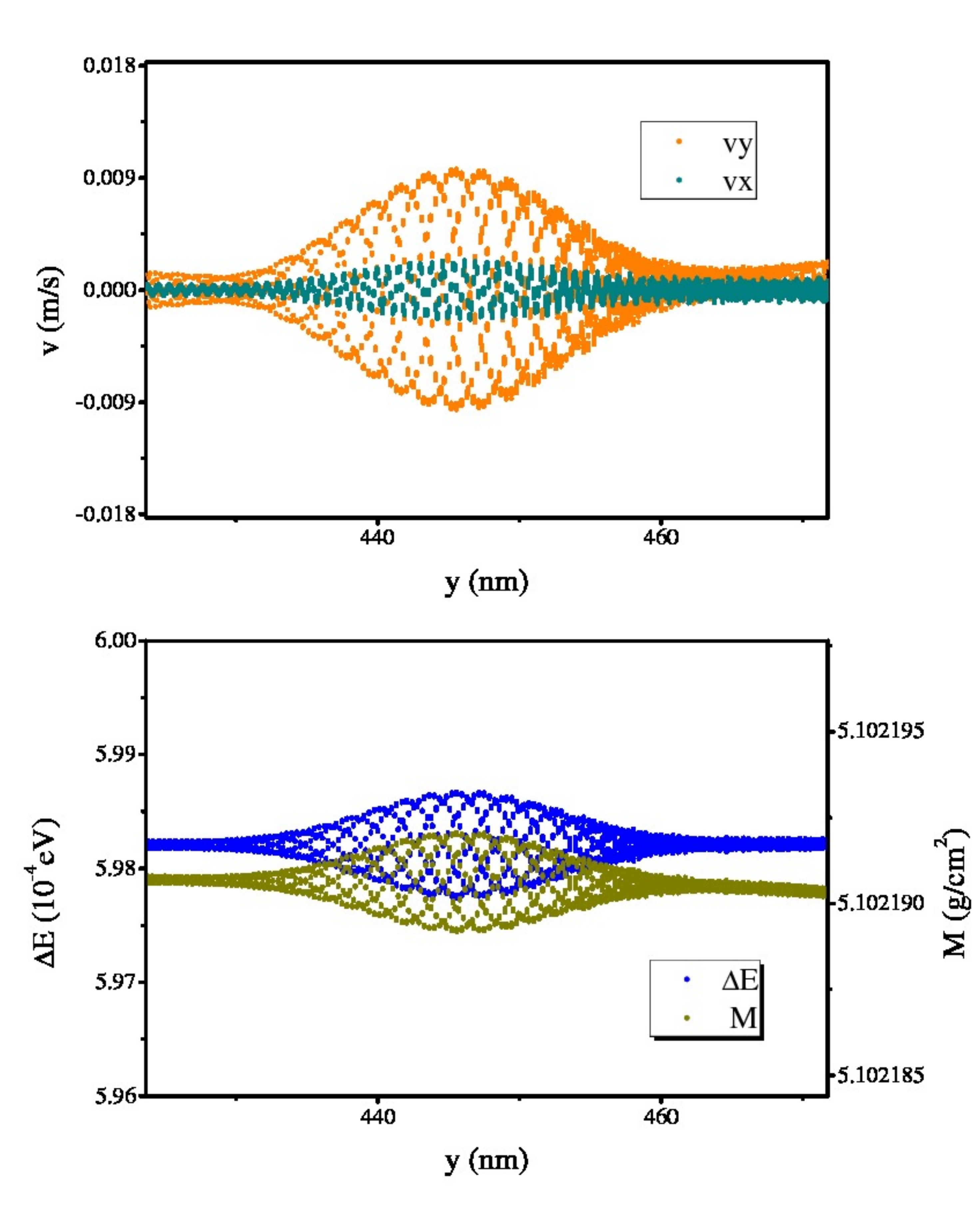}\caption{(Color online) The breather-like soliton is investigated. Above it
shows the longitudinal and transverse velocity distributions of the
breather-like soliton. Down it shows the energy and reduced density
distributions of the breather-like soliton.}

\end{figure}
\begin{figure}
\includegraphics[scale=0.2]{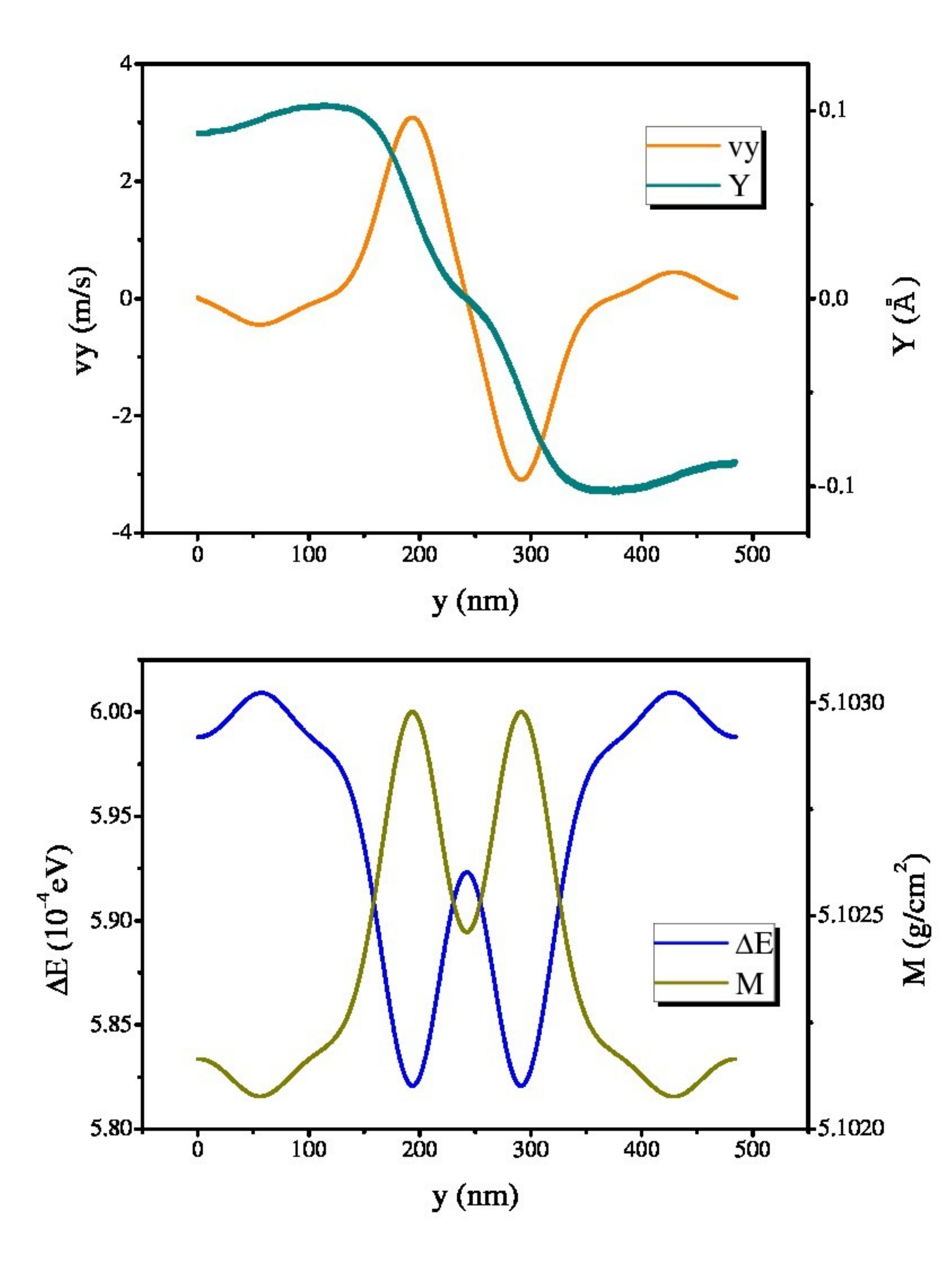}\caption{(Color online) Using different initial conditions, another two small
solitons are identified in Graphene $\theta(0)$ They are also longitudinal
solitons. }

\end{figure}

Typical solitons in Graphene $\theta(30)$ are shown in Fig. 5 and
Fig. 6. A pair of transverse solitons and a pair of longitudinal solitons
are identified. 

The velocity and displacement distributions of the transverse and
longitudinal solitons are shown in Fig. 5. The displacement distributions
of the longitudinal solitons are not easy to distinguish from the
background fluctuation. However their shapes resemble the longitudinal
solitons in Graphene $\theta(0)$ as well. Their energy and density
distributions are shown in Fig. 6. The transverse solitons aggregate
energy and lack density. The longitudinal solitons lack energy towards
the background. When two transverse solitons collide together we can
identify them lacking of density then. %
\begin{figure}
\includegraphics[scale=0.2]{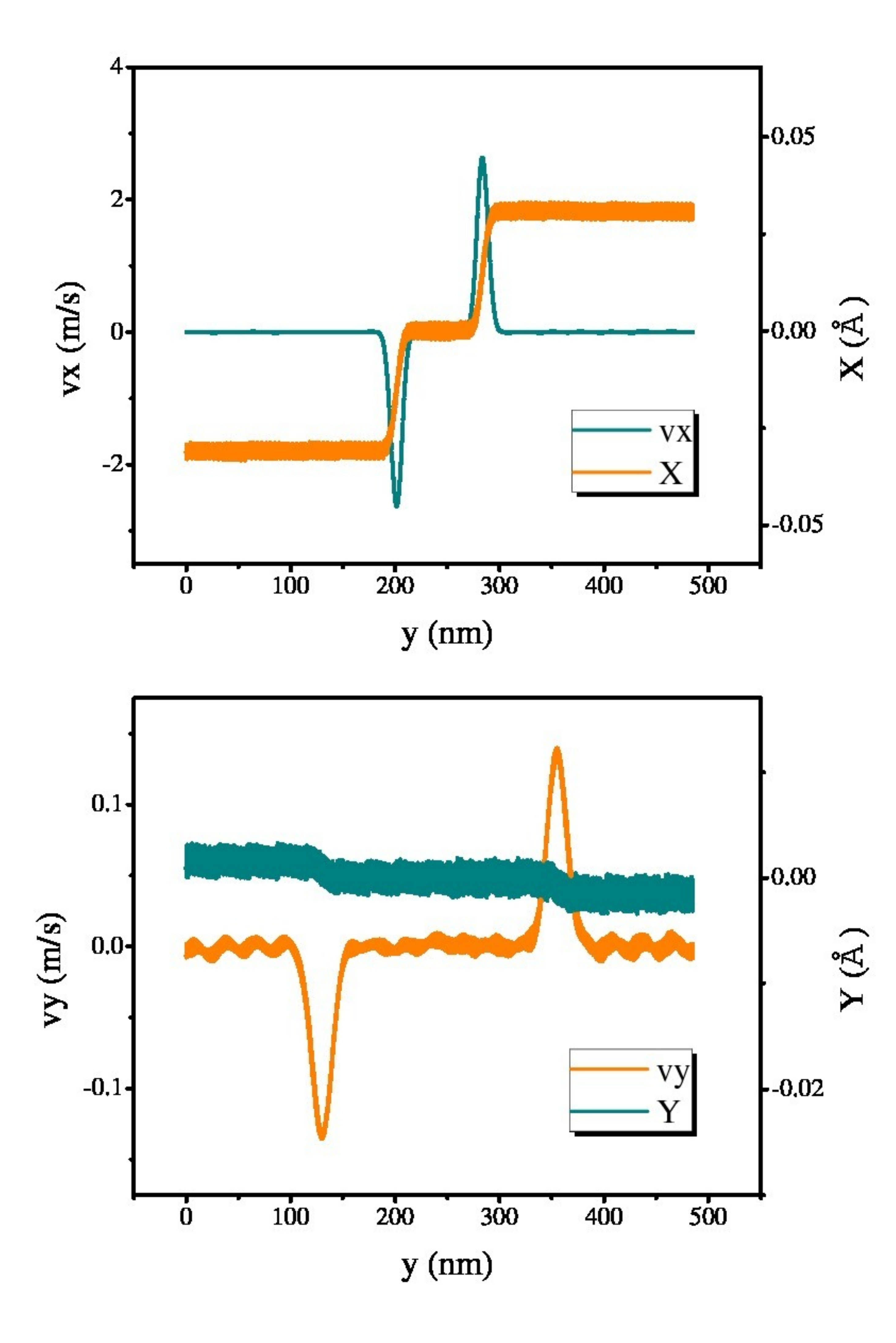}\caption{(Color online) Typical solitons in Graphene $\theta(30)$ are investigated.
Above it shows the velocity and displacement distributions of the
transverse solitons. Down it shows the velocity and displacement distributions
of the longitudinal solitons. }

\end{figure}

Due to the momentum conservation solitons emerge in pairs with opposite
velocity localization value. Since the graphene flakes have no compression
or stretching the total displacements are also zero. The lengths of
some solitons are comparable with the mean free path of phonons\citet{27.mean free path}.
It indicates a long length might be needed to consider the contribution
of solitons in thermal conduction. We only observe random fluctuations
in out-of-plane distribution. Such difference might affect thermal
conduction when flexural modes are enhanced or suppressed\citet{28.supported and suspended}.

In order to confirm the wave profiles are solitons we have to study
their propagation and collision. Solitons shall preserve their identities
in propagation and after collision. It is the usual way to identify
soliton nature in nonlinear kinetics. It is shown in Fig. 7-10 how
solitons propagate and collide in the graphene flakes. The propagation
and collision of the longitudinal and breather-like solitons in Graphene
$\theta(0)$ are shown in Fig. 7 and Fig. 8. The propagation and collision
of the transverse and longitudinal solitons in Graphene $\theta(30)$
are shown in Fig. 9 and Fig. 10. %
\begin{figure}
\includegraphics[scale=0.2]{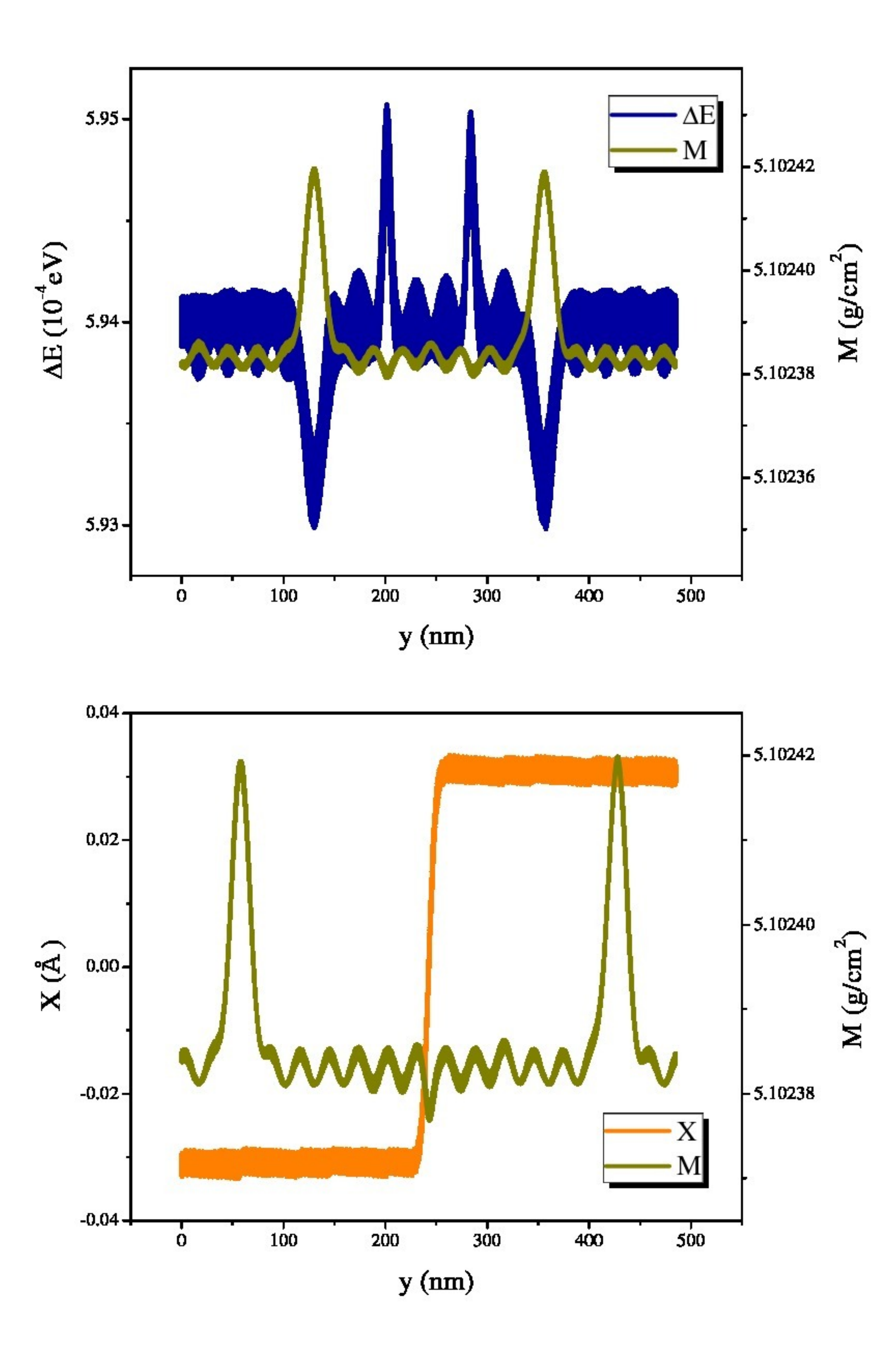}\caption{(Color online) Typical solitons in Graphene $\theta(30)$ are investigated.
Above it shows the energy and density distributions of the transverse
and longitudinal solitons. Down it shows the energy and density distributions
when the two longitudinal soliton collide together.}

\end{figure}
\begin{figure}
\includegraphics[scale=0.3]{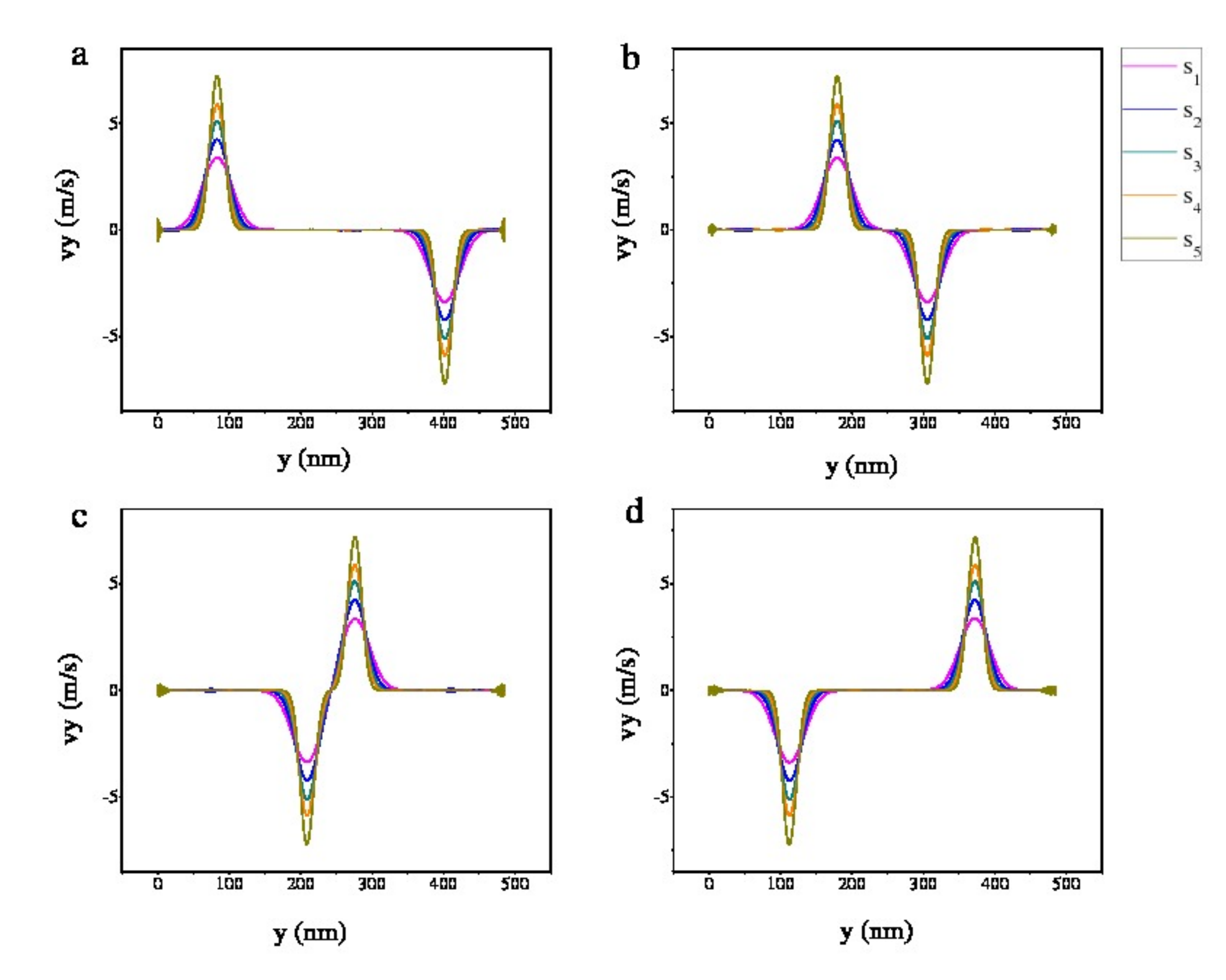}\caption{(Color online) Propagation and collision of the two longitudinal solitons
in Graphene $\theta(0)$ with different amplitudes. The time interval
between two neighboring events is 4.8 ps from (a) to (d). }

\end{figure}
\begin{figure}
\includegraphics[scale=0.2]{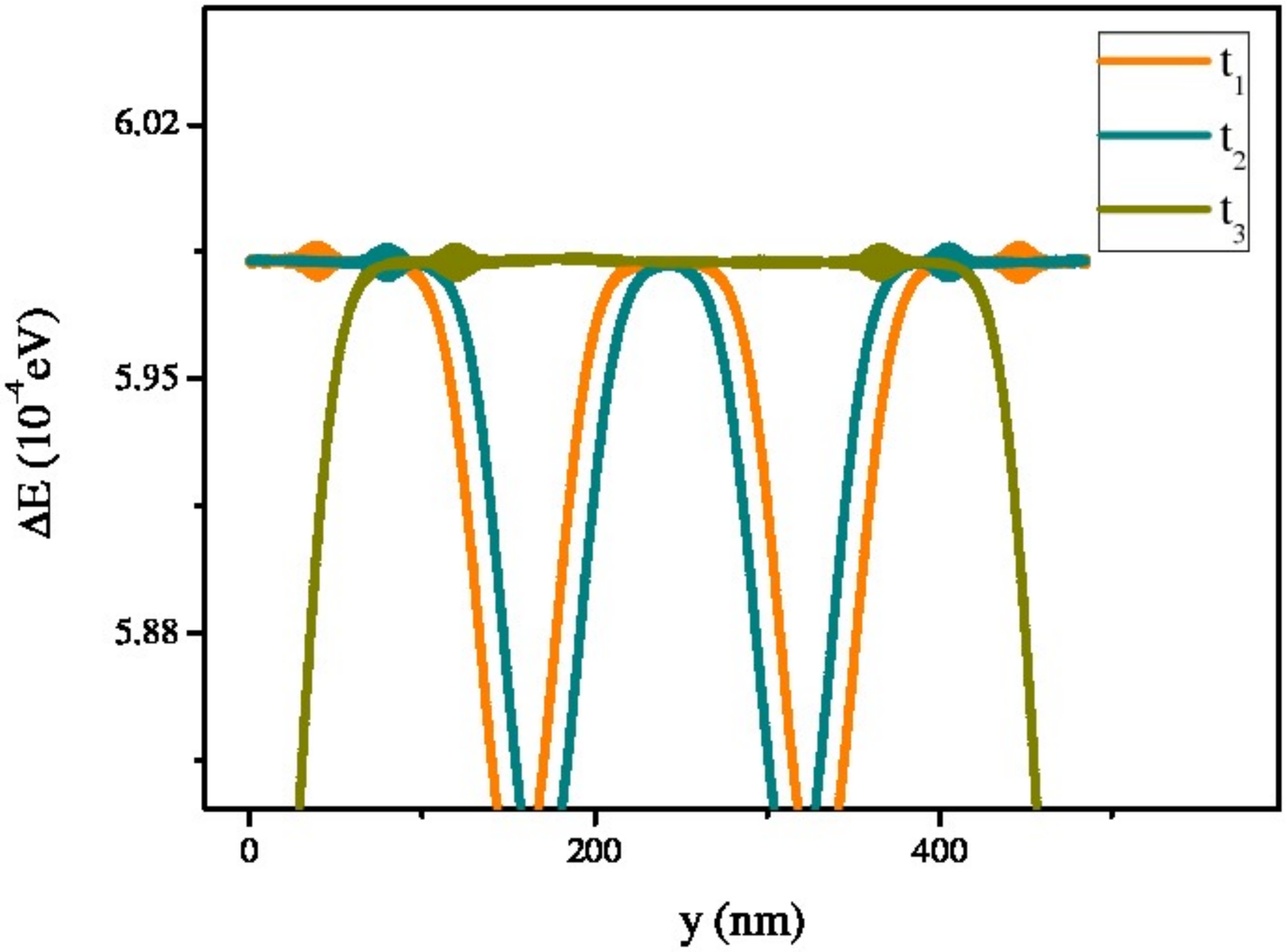}\caption{(Color online) Propagation of the breather-like solitons in Graphene
$\theta(0)$. The time interval is 8 ps between two neighboring events.}

\end{figure}

It is shown in Fig. 7 the propagation and collision of the longitudinal
solitons with different amplitudes in Graphene $\theta(0)$. Their
propagating velocity is 20.0 km/s and it is smaller than the longitudinal
sound speed which is 21.3 km/s\citet{29.sound speed}. The propagation
velocity is close to the sound speed. The propagating velocity seems
insensitive to the amplitudes. Their collisions are elastic without
changing their identities. It confirms they are solitons. 

It is shown in Fig. 8 the propagation of the breather-like solitons
in Graphene $\theta(0)$. Their propagating velocity is 5.0 km/s.
Their collisions are also elastic. 

It is shown in Fig. 9 the propagation and collision of the transverse
solitons in Graphene $\theta(30)$. It is shown in Fig. 10 the propagation
and collision of the longitudinal solitons in Graphene $\theta(30)$.
The propagating velocity of transverse solitons is 11.4 km/s. The
propagating velocity of the longitudinal solitons is 20.0 km/s. Both
velocities are smaller than the relative sound speed which is 21.3
km/s (longitudinal sound speed) and 13.6 km/s (transverse sound speed)\citet{29.sound speed}.
Their propagating velocities are also weakly dependent upon the amplitudes.
Their collisions are elastic as well.%
\begin{figure}
\includegraphics[scale=0.3]{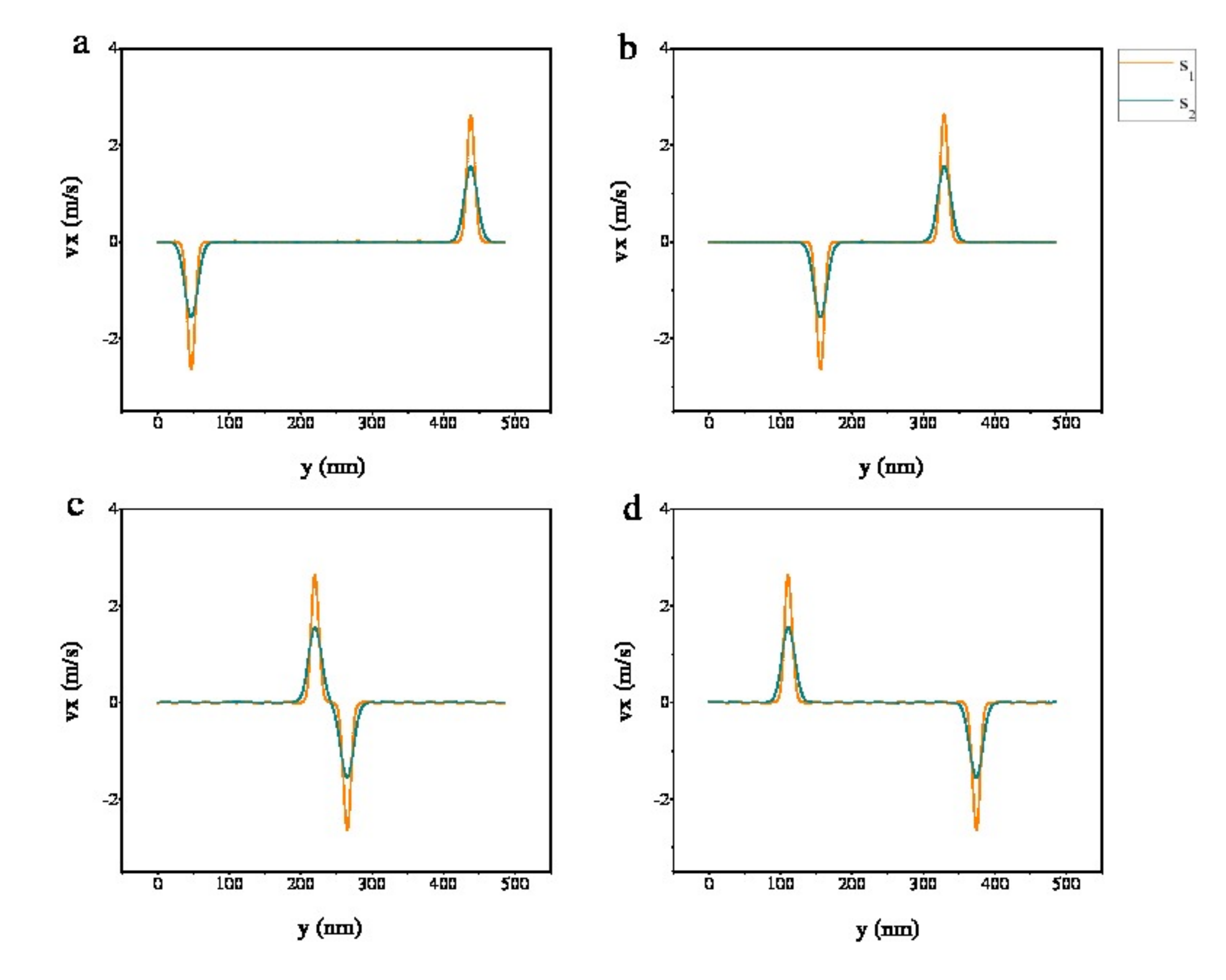}\caption{(Color online) Propagation and collision of the transverse solitons
in Graphene $\theta(30)$ with different amplitudes. The time interval
between two neighboring events is 9.6 ps from (a) to (d).}

\end{figure}
\begin{figure}
\includegraphics[scale=0.3]{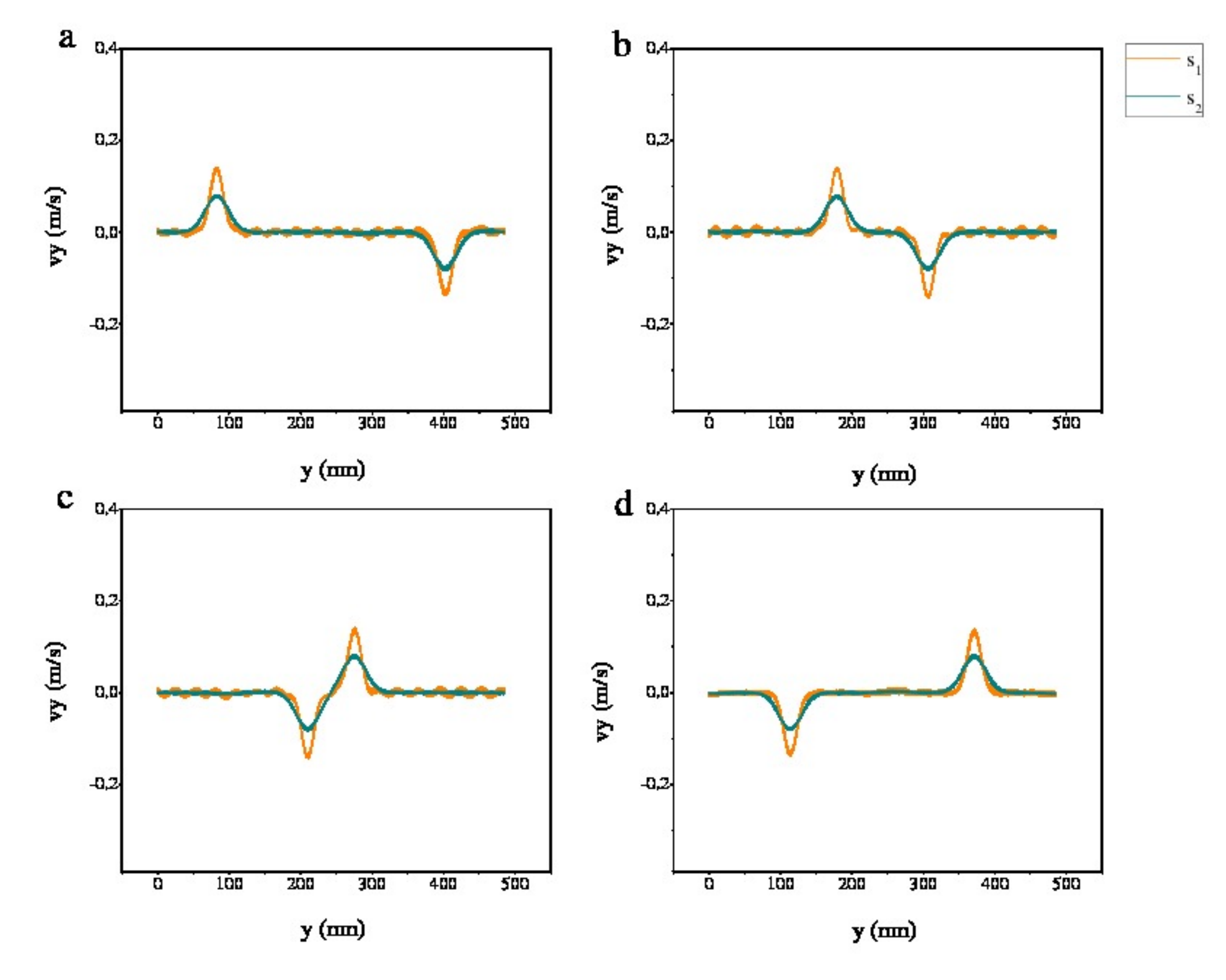}\caption{(Color online) Propagation and collision of the longitudinal solitons
in Graphene $\theta(30)$ with different amplitudes. The time interval
between two neighboring events is 4.8 ps from (a) to (d).}

\end{figure}

Solitons shall have interactions (decelerations and accelerations)
in collisions due to nonlinearity. It is distinctively different from
linear waves which have no interactions due to superposition principle.
Solitons would deviate from their original trajectories due to collisions.
They are not in the position which would be anticipated by extending
their original trajectories. It means solitons are phase-shifted.
Phase shift is a well known behavior of solitons in collisions\citet{30.book,31.taojin}.

It is shown in Fig. 11 the spatial-temporal trajectories of the longitudinal
solitons in Graphene $\theta(0)$. Straight lines are observed before
or after collision. In collision, solitons are first decelerated and
then accelerated by a spatial jump. The average propagating velocity
is 22.3 km/s during collision thus solitons exhibit an averaged acceleration
effect. The dash lines represent the original trajectories of the
solitons assuming there was no collision. It clearly illustrates the
averaged acceleration effect brings forth the phase shift of solitons.
Spatial jumps are also the important behavior of solitons in collision\citet{31.taojin}.
It is shown in Fig. 11 how the spatial jump happens. Since we identify
the position of solitons according to the extreme value points. So
it is difficult to do so in some critical moments when there are four
extreme value points. It behaves like a spatial jump if we neglect
those moments. 

It is shown in Fig. 12 the spatial-temporal trajectories of two transverse
and two longitudinal solitons in Graphene $\theta(30)$. The transverse
solitons are first decelerated and then accelerated by a spatial jump
in collision. The averaged propagating velocity is 13.9 km/s during
collision. It exhibits an averaged acceleration effect in collision.
The phase shift of the transverse solitons is clearly illustrated
by the dash lines. Similar behaviors are also observed in collision
between two longitudinal solitons. The averaged propagating velocity
is 22.3 km/s in collision during collision. The phase shift of the
longitudinal solitons is also clearly illustrated by the dash lines. 

It is shown in Fig. 13 the spatial-temporal trajectories of a transverse
soliton and a longitudinal soliton in Graphene $\theta(30)$. It describes
the collision between a transverse soliton and a longitudinal soliton.
Unlike collision between two transverse solitons or two longitudinal
solitons, they would simply pass through each other without any interactions.
Thus only straight lines are observed in their trajectories.%
\begin{figure}
\includegraphics[scale=0.2]{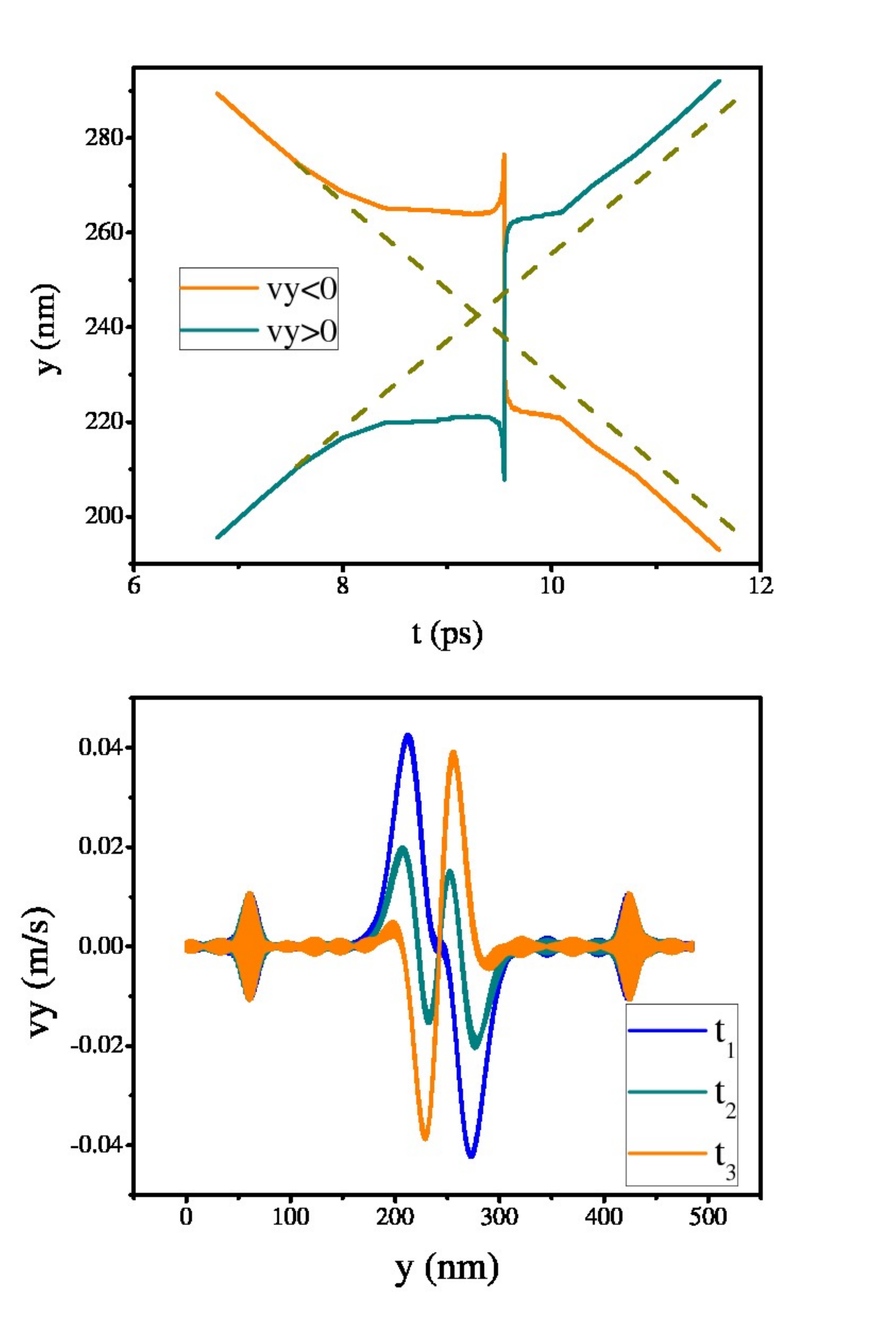}\caption{(Color online) Above it shows the spatial-temporal trajectories of
two longitudinal solitons in Graphene $\theta(0)$. Dash lines are
added as auxiliary lines to illustrate the original trajectories of
the solitons assuming there was no collision. Down it shows the velocity
distributions around the spatial jump. The time interval between the
two neighboring events is 0.0075 ps. $t_{2}$ is the critical moment
in the spatial jump.}

\end{figure}
\begin{figure}
\includegraphics[scale=0.2]{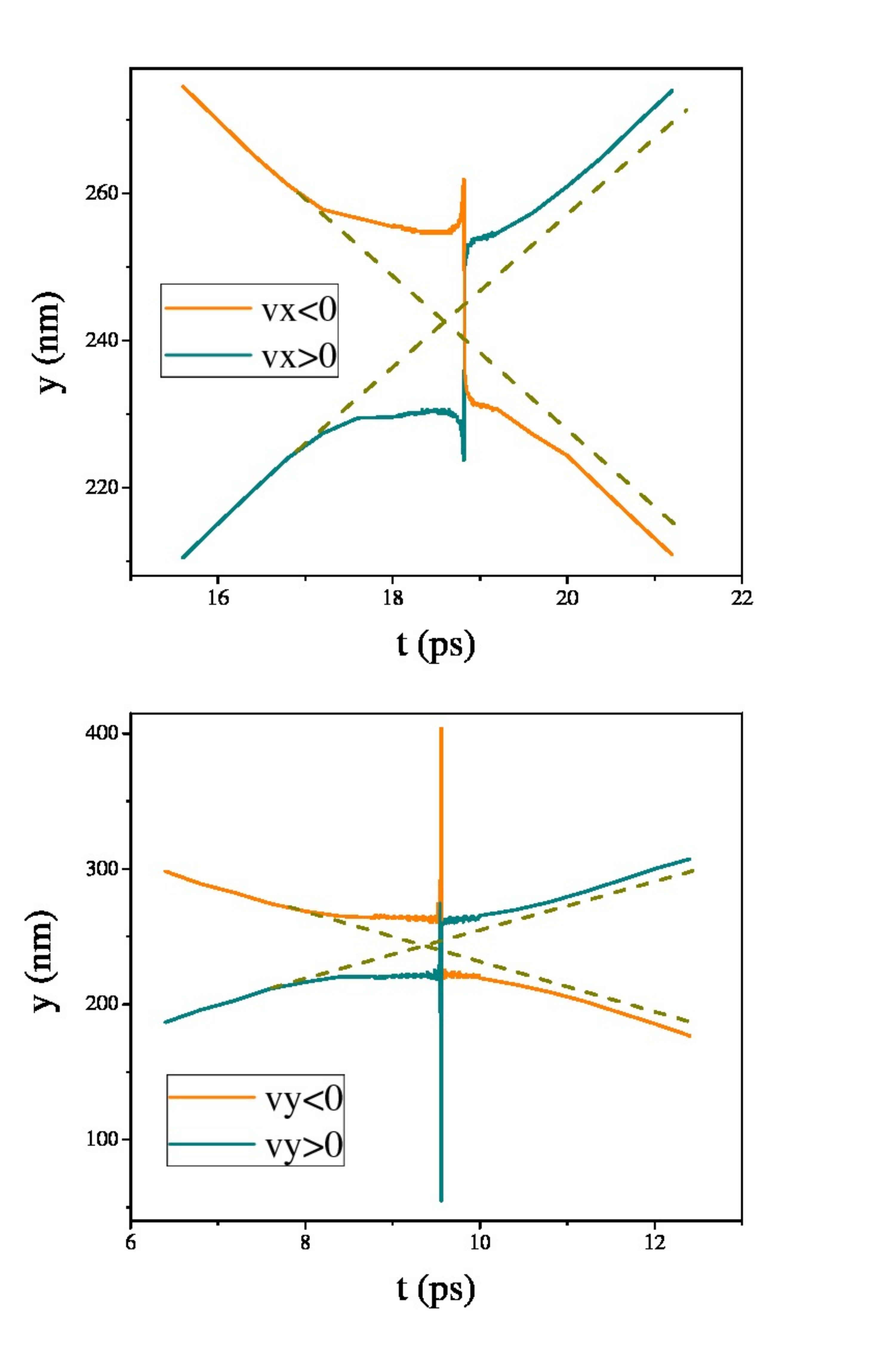}\caption{(Color online) Above it shows the spatial-temporal trajectories of
two transverse solitons in Graphene $\theta(30)$. Down it shows the
spatial-temporal trajectories of two longitudinal solitons in Graphene
$\theta(30)$. Dash lines are added as auxiliary lines to illustrate
the original trajectories of the solitons assuming there was no collision.}

\end{figure}
\begin{figure}
\includegraphics[scale=0.2]{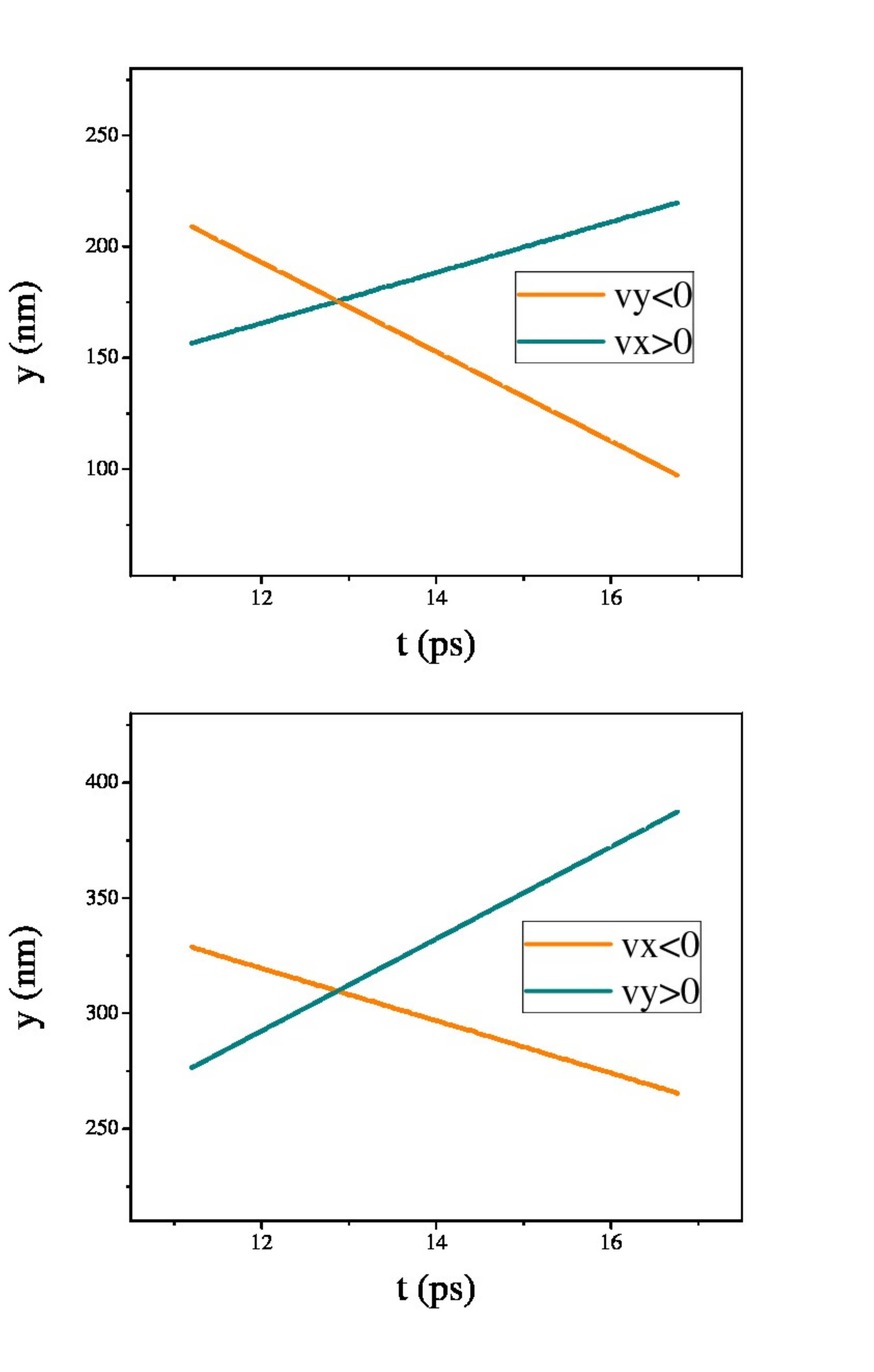}

\caption{(Color online) Above it shows the spatial-temporal trajectories of
a transverse soliton propagate forward and collide with a longitudinal
soliton in Graphene $\theta(30)$. Down it shows the spatial-temporal
trajectories of a transverse soliton propagate backward and collide
with a longitudinal soliton in Graphene $\theta(30)$.}

\end{figure}

Using different initial conditions and cooling time steps even more
solitons might emerge in graphene with similar kinetic properties.
It is shown in Fig. 14 and Fig. 15. %
\begin{figure}
\includegraphics[scale=0.2]{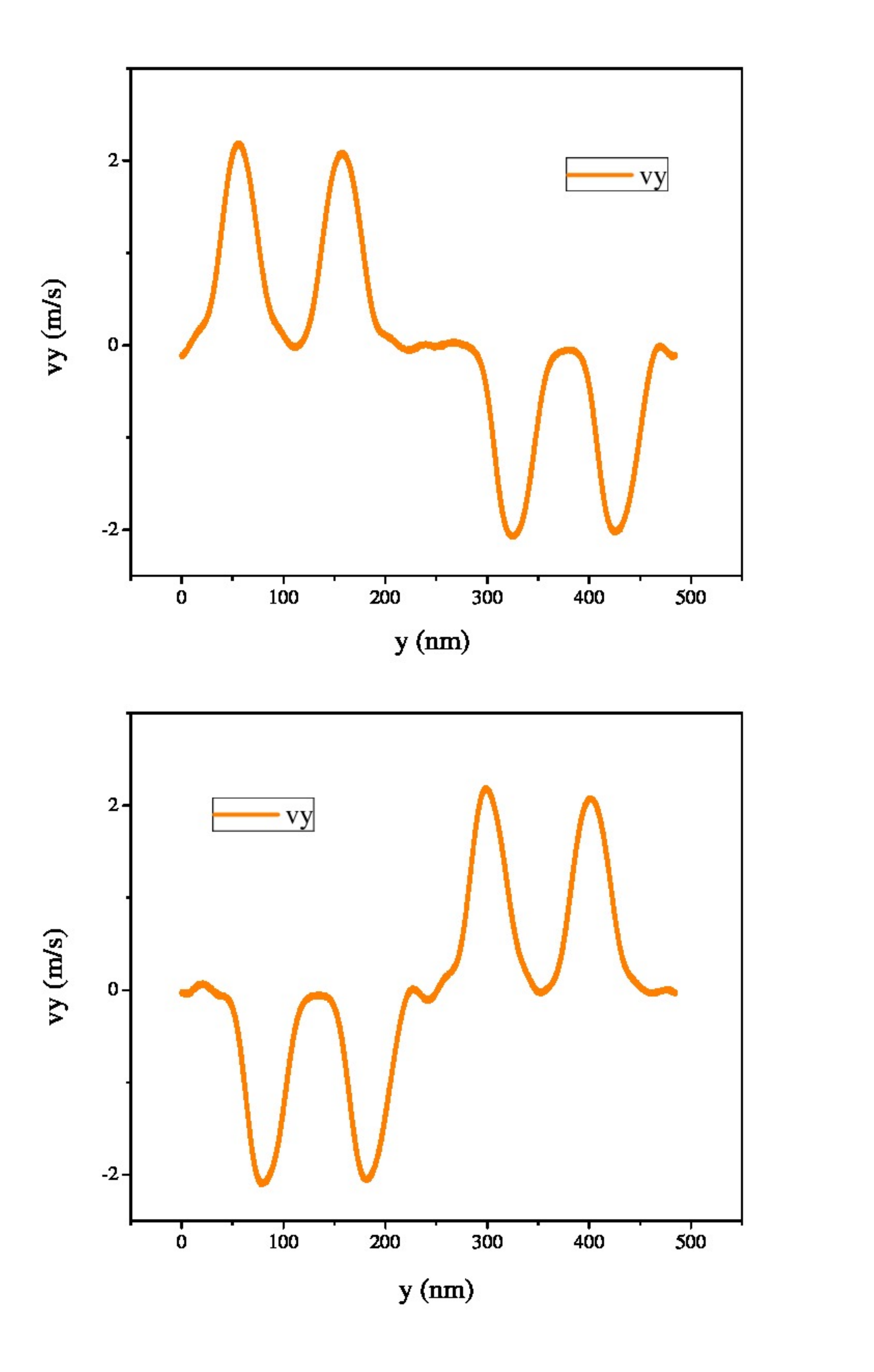}\caption{(Color online) Four solitons in Graphene $\theta(0)$. The interval
between the two neighboring events is 12 ps.}

\end{figure}
\begin{figure}
\includegraphics[scale=0.2]{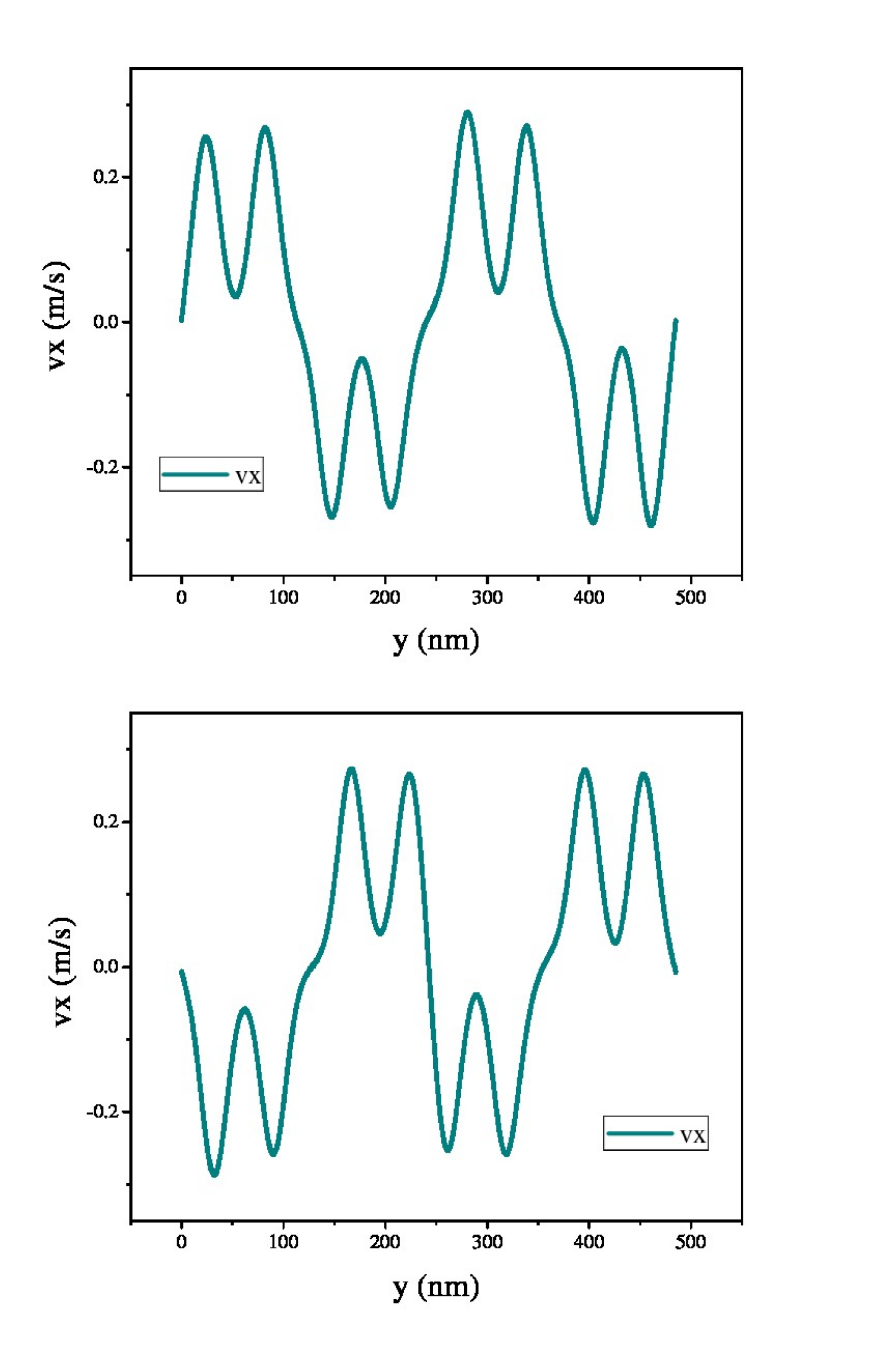}\caption{(Color online) Eight solitons in Graphene $\theta(30)$. The interval
between the two neighboring events is 10 ps. }

\end{figure}

The emergence of solitons is observed by cooling procedures. It is
quite different from the usual way in which solitons are generated
by imposing a strong impulse upon the lattice\citet{31.taojin,32.kick}.
Thus a different theoretical approach is required to obtain the soliton
solution. Here we use the multiscale asymptotic method to derive a
NLS equation to describe the nonlinear process in graphene. Multiscale
asymptotic method first developed in 1969\citet{33.original multiscale}
assumes all higher-order harmonics are small enough so that the fundamental
frequencies determine the kinetic behaviors of the system. It is widely
used in different nonlinear systems to obtain soliton solutions\citet{34.multiscale 01,35.multiscale 02.breather,36.FPU multiscale 03}. 

We use the Brenner\textquoteright{}s potential\citet{25.Brenner OR}
to describe the interactions between carbon atoms in the analytical
model here. It is the 1st simplification in our analytical model.

Here we take Graphene $\theta(0)$ as an example to obtain the soliton
solution. Following similar steps\citet{19.nanotube KDV,20.supersonic KDV graphite}
we expand the interatomic potential into Taylor series up to fourth
order around the equilibrium position in Graphene $\theta(0)$. For
simplicity only the longitudinal displacement Y is considered. It
is the 2nd simplification in our analytical model.

The Hamiltonian is $H=H_{0}+\sum_{n=1}^{N_{l}}H_{n}$ where $H_{n}$
is the perturbation energy in $n$th layer. Thus $H_{n}=m_{n}H_{n}^{a}$
where $m_{n}$ is the number of atoms in $n$th layer and we get

$\,$

$H_{n}^{a}=\frac{b}{2}(Y_{n+1}-Y_{n})^{2}-\frac{bp}{3}(Y_{n+1}-Y_{n})^{3}+\frac{bq}{4}(Y_{n+1}-Y_{n})^{4}$
$\,$(3)

$\,$

Here $Y_{n}=y_{n}-Y_{n0}$ is the longitudinal displacements of atoms
around their equilibrium position in $n$th layer and $b=58.93eV$,
$p=2.796$, $q=1.990$. 

Then we can get the kinetic equation to describe the longitudinal
motion as:

$\;$

$m\frac{d^{2}Y_{n}}{d^{2}t}=\frac{b}{l_{0}^{2}}[Y_{n+1}-2Y_{n}+Y_{n-1}]-\frac{bp}{l_{0}^{2}}[(Y_{n+1}-Y_{n})^{2}-(Y_{n}-Y_{n-1})^{2}]+\frac{bq}{l_{0}^{2}}[(Y_{n+1}-Y_{n})^{3}-(Y_{n}-Y_{n-1})^{3}]$
$\,$$\,\qquad\qquad\qquad$(4)

$\;$

Here $m=1.99\times10^{-26}$kg and $l_{0}=1.212\textrm{\AA}$ which
is the equilibrium distance between two layers. 

We then expand the displacement by small excitations up to first-order
perturbation term

$\,$

$Y_{n}=\sum_{\nu=1}^{\infty}\varepsilon^{\nu}u_{n}^{(\nu)}(\tau,\xi_{n},\phi_{n})\thickapprox\varepsilon u_{n}^{(1)}(\tau,\xi_{n},\phi_{n})$
$\,$$\qquad\qquad$(5)

$\,$

Here $u_{n}^{(1)}=\sum_{l=-\infty}^{\infty}u_{n,l}^{(1)}(\tau,\xi_{n})exp(il\phi_{n})$
is expanded in harmonic terms with $\tau=\sqrt{\frac{b}{ml_{0}^{2}}}\varepsilon^{2}t$,
$\xi_{n}=\varepsilon nl_{0}-\lambda\tau/\varepsilon$, $\phi_{n}=knl_{0}-\omega\tau/\varepsilon^{2}$. 

Here we simplify the expansions again by neglecting the high-order
harmonics $u_{n,l}^{(1)}(\tau,\xi_{n})=0$, if $\left|l\right|>1$.
It is the 3rd simplification in our analytical model.

Substituting $Y_{n}=\varepsilon u_{n,0}^{(1)}(\tau,\xi_{n})+\epsilon u_{n,1}^{(1)}(\tau,\xi_{n})exp(i\phi_{n})+\varepsilon u_{n,1}^{(1)*}(\tau,\xi_{n})exp(-i\phi_{n})$
into the kinetic equation and equating the coefficients of $\varepsilon$.
After that we derive the following equation:

$\,$

$i\frac{\partial u_{n,1}^{(1)}}{\partial\tau}=\frac{\omega l_{0}^{2}}{8}\frac{\partial^{2}u_{n,1}^{(1)}}{\partial\xi_{n}^{2}}+[\frac{3}{2}q\omega^{3}-\omega p^{2}(4+\omega^{2})]\left|u_{n.1}^{(1)}\right|^{2}u_{n,1}^{(0)}+8\omega Cp^{2}u_{n,1}^{(1)}$
$\qquad\;\qquad\qquad\qquad\qquad\qquad\qquad\;\qquad\:\qquad$(6)

$\,$

Here $C$ is an integration constant. By substituting variables it
is reduced to a standard cubic NLS equation:

$\,$

$i\frac{\partial\psi}{\partial\eta}-\frac{\sigma}{2}\frac{\partial^{2}\psi}{\partial\delta^{2}}+\left|\psi\right|^{2}\psi=0$
$\qquad\qquad\qquad\qquad\:\:\;\quad\qquad$(7)

$\,$

Here $\psi=exp(i8p^{2}\omega\int Cd\tau)u_{n,1}^{(1)}$, $\sigma=\frac{1}{(p^{2}-\frac{3}{2}q)\omega^{2}+4p^{2}}$,
$\eta=\omega\tau/\sigma$, $\delta=\frac{2}{l_{0}}\xi_{n}$. Here
we have $\sigma=\frac{1}{4.83\omega^{2}+31.3}>0$ which means it is
a normal NLS equation. %
\begin{figure}
\includegraphics[scale=0.2]{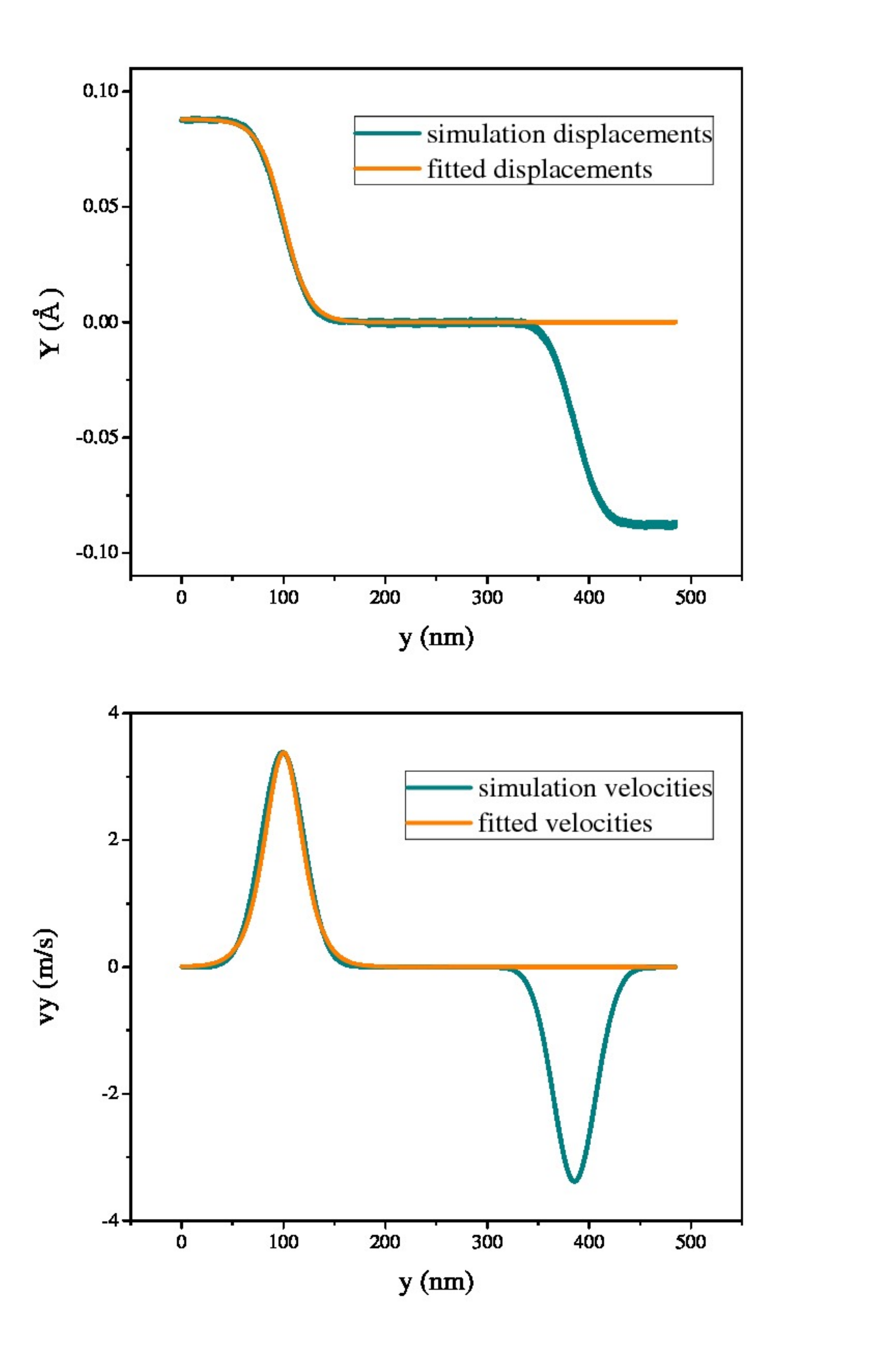}\caption{(Color online) Comparing the analytical soliton solution with the
soliton obtained in simulations. Above it shows the displacement distribution
agrees with the simulation results. Down it shows the velocity distribution
also agrees with the simulation results.}

\end{figure}

A soliton solution is possible in the NLS equation above:

$\,$

$Y_{n}=A[cosX-2p\sqrt{\sigma}]tanh\Theta-Atan\varphi sinX+A'$ $\qquad\;\;$(8)

$\,$

Here $A'$ is an integration constant and $A$, $\alpha$, $\varphi$
are free parameters. $v_{1}$ is the propagation velocity of the soliton.

$\,$

$X=knl_{0}+2\alpha-[2\alpha cos\frac{kl_{0}}{2}+(2-\alpha^{2}-\frac{A^{2}}{2\sigma cos^{2}\varphi}+\frac{4p^{2}A^{2}}{cos^{2}\varphi})sin\frac{kl_{0}}{2}]\sqrt{\frac{b}{ml_{0}^{2}}}t$

$\,$

$\Theta=\frac{A}{\sqrt{\sigma}}(nl_{0}-tv_{1})$

$\,$

$v_{1}=[cos\frac{kl_{0}}{2}-(\alpha+\frac{Atan\varphi}{2\sqrt{\sigma}})sin\frac{kl_{0}}{2}]\sqrt{\frac{b}{m}}$

$\,$

$y_{n}=Y_{n}+Y_{n0}=Y_{n}+nl_{0}$

$\,$

The complete form of the soliton solution is complicated. It can be
further simplified by considering the kinetic properties of solitons
in the molecular simulations. From the previous study, we know the
velocities of the solitons are insensitive to the amplitudes. Thus
the amplitude dependent parameter in $v_{1}$ might be neglected.
It means $tan\varphi=0$ here. It is the 4th simplification in our
analytical model.

The propagating velocities of solitons are close to the relative sound
speed. It also can be used as a valid simplification. The relative
sound speed is $v_{2}=\sqrt{\frac{b}{m}}$. It means $v_{1}\cong v_{2}$.
It is the 5th simplification in our analytical model.

From the 5th simplification we can obtain

$\,$

$[cos\frac{kl_{0}}{2}-(\alpha+\frac{Atan\varphi}{2\sqrt{\sigma}})sin\frac{kl_{0}}{2}]\sqrt{\frac{b}{m}}\cong\sqrt{\frac{b}{m}}$
$\qquad\qquad\quad\quad$(9)

$\,$

Taking the 4th simplification into account, we have

$\,$

$cos\frac{kl_{0}}{2}-\alpha sin\frac{kl_{0}}{2}\cong1$ $\qquad\qquad\qquad\qquad\quad\qquad\qquad$
(10)

$\,$

So we may obtain the following relation that:

$\,$

$cos\frac{kl_{0}}{2}\cong1$, $sin\frac{kl_{0}}{2}\cong0$, $\alpha\ll1$
$\qquad\qquad\qquad\qquad\quad\quad$(11) 

$\,$

Taking $tan\varphi=0$, $cos\frac{kl_{0}}{2}\cong1$, $sin\frac{kl_{0}}{2}\cong0$
and $\alpha\ll1$ into Eq. 8, we finally can obtain the soliton solution
in the continuum approach as:

$\,$

$Y=A_{0}tanh[c(y-tv_{1})]+A'$ $\qquad\qquad\qquad\qquad\:\:\qquad$(12)

$\,$

Here $A_{0}$ and $c$ are determined by the anharmonic parameters
$p$ and $q$ in Eq. 3.

$\,$

$A_{0}=A(\sqrt{1-4\alpha^{2}}-2p\sqrt{\sigma})=A(\sqrt{1-4\alpha^{2}}-\sqrt{\frac{4p^{2}}{(p^{2}-\frac{3}{2}q)\omega^{2}+4p^{2}}})$
$\qquad\qquad\qquad\qquad\qquad\qquad\quad\qquad\;$(13)

$\,$

$c=A/\sqrt{\sigma}=A\sqrt{(p^{2}-\frac{3}{2}q)\omega^{2}+4p^{2}}$
$\qquad\qquad\qquad\;$ (14)

$\,$

The relative longitudinal velocity distribution of the soliton solution
is:

$\,$

$vy=\frac{dY}{dt}=-A_{0}cv_{1}sech^{2}[c(y-tv_{1})]$ $\qquad\qquad\:\:\:\:\qquad\;$
(15)

$\,$

In a given time $t_{0}$ the relative soliton solution should be:

$\,$

$Y=A_{0}tanh[c(y-y_{0})]+A'$ $\qquad\qquad\qquad\qquad\qquad\;$
(16)

$\,$

$vy=-A_{0}cv_{1}sech^{2}[c(y-y_{0})]$ $\quad\qquad\qquad\qquad\:\qquad\;$
(17)

$\,$

To investigate the validity of our analytical results, we first try
to fit the soliton obtained in the simulation with the soliton solution
we derived. Next we try to use the fitted soliton solutions as the
initial conditions to investigate the propagation and collision of
the fitted solitons.

We first fit the displacement distribution to determine $A_{0}$,
$c$ and $A'$. When the parameters are obtained we use them to fit
the velocity distribution. The fitted displacement and velocity distributions
($v_{1}=20.0$ km/s) are:

$\,$

$Y=-0.044tanh[0.04(y-100)]+0.044$ $\qquad\;\;\;\;\;\qquad$ (18)

$\,$

$vy=3.5sech^{2}[0.04(y-100)]$ $\qquad\qquad\qquad\qquad\qquad$ (19)

$\,$

It is shown in Fig. 16 the results fit the left side soliton well.
The right side soliton can also be well fitted in the same way. Next
we use the fitted solitons as the initial conditions to investigate
their propagation and collision properties. It is shown in Fig. 17
that the propagation and collision properties of the fitted solitons
are almost the same as the solitons obtained in the previous molecular
dynamics simulations. It states the derived soliton solution can describe
the solitons obtained in the molecular dynamics simulations well .%
\begin{figure}
\includegraphics[scale=0.3]{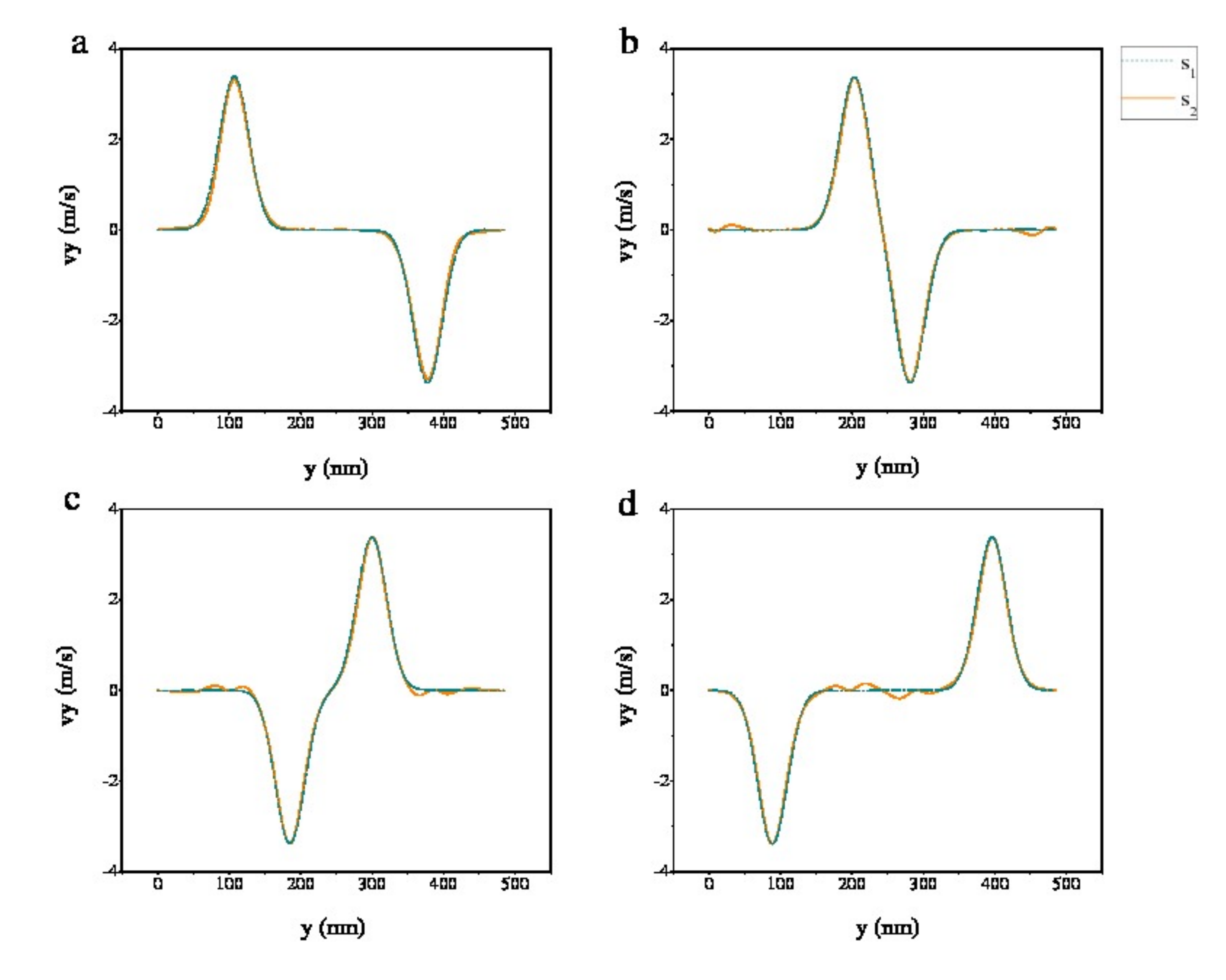}\caption{(Color online) $s_{1}$ is the solitons obtained in the previous molecular
dynamics simulations by cooling procedures. $s_{2}$ is the solitons
derived in the analytical results. The propagation and collision properties
resemble each other well.}

\end{figure}
\begin{figure}
\includegraphics[scale=0.3]{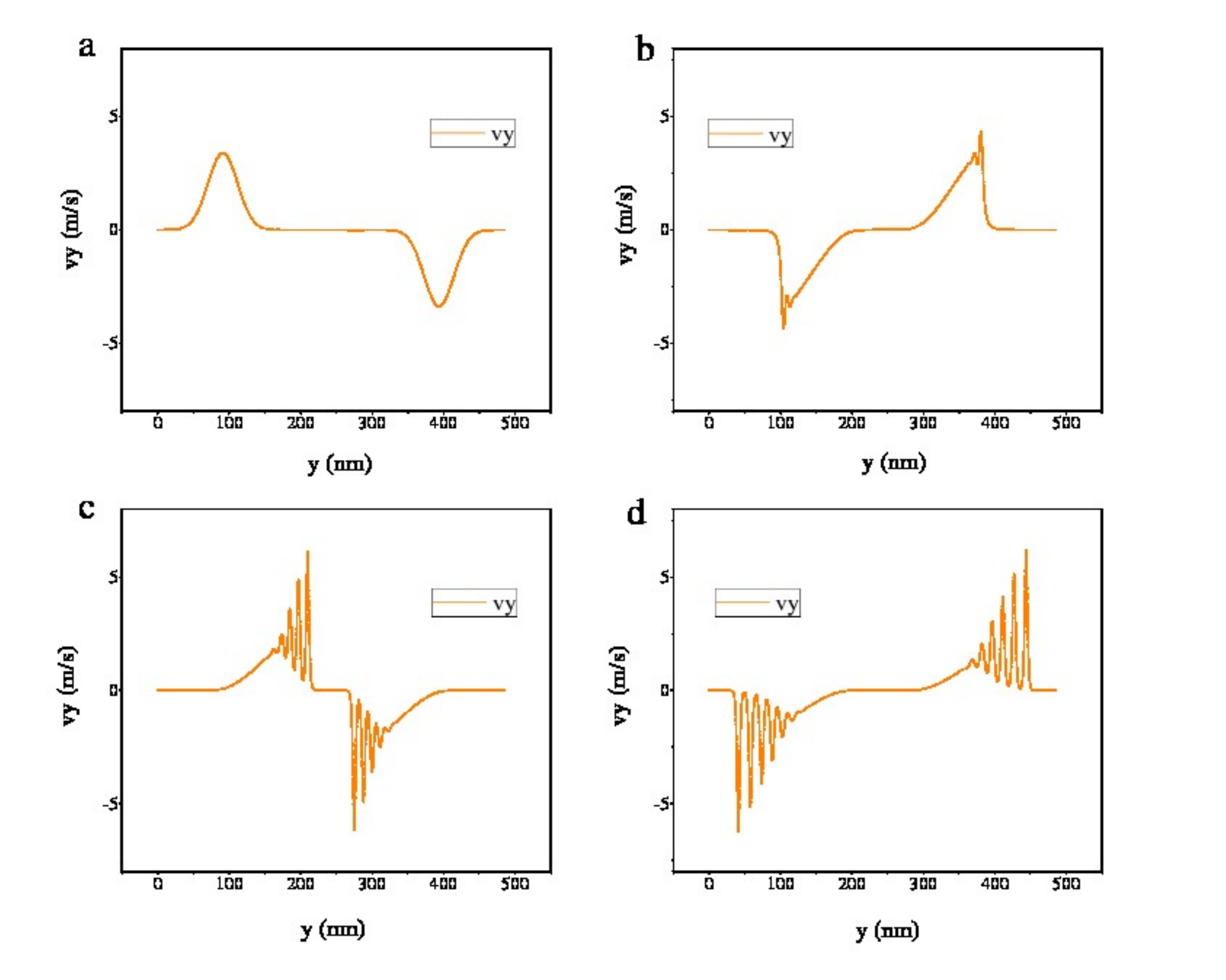}\caption{(Color online) Long time behavior of the longitudinal solitons in
Graphene $\theta(0)$. The time interval between two neighboring events
is 1.2 ns from (a) to (d).}

\end{figure}
\begin{figure}
\includegraphics[scale=0.3]{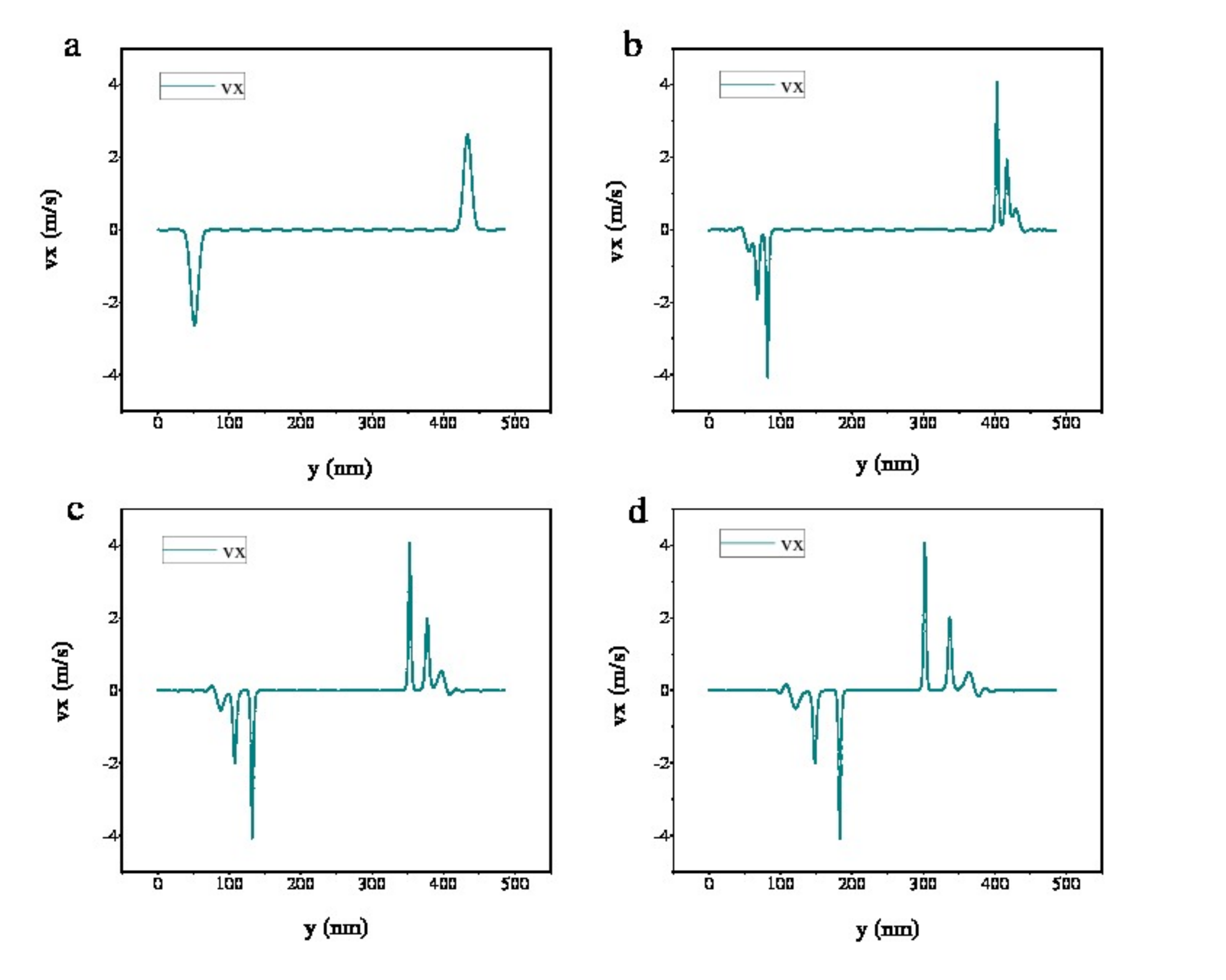}\caption{(Color online) Long time behavior of the transverse solitons in Graphene
$\theta(30)$. The time interval between two neighboring events is
1.2 ns from (a) to (d).}

\end{figure}
\begin{figure}
\includegraphics[scale=0.2]{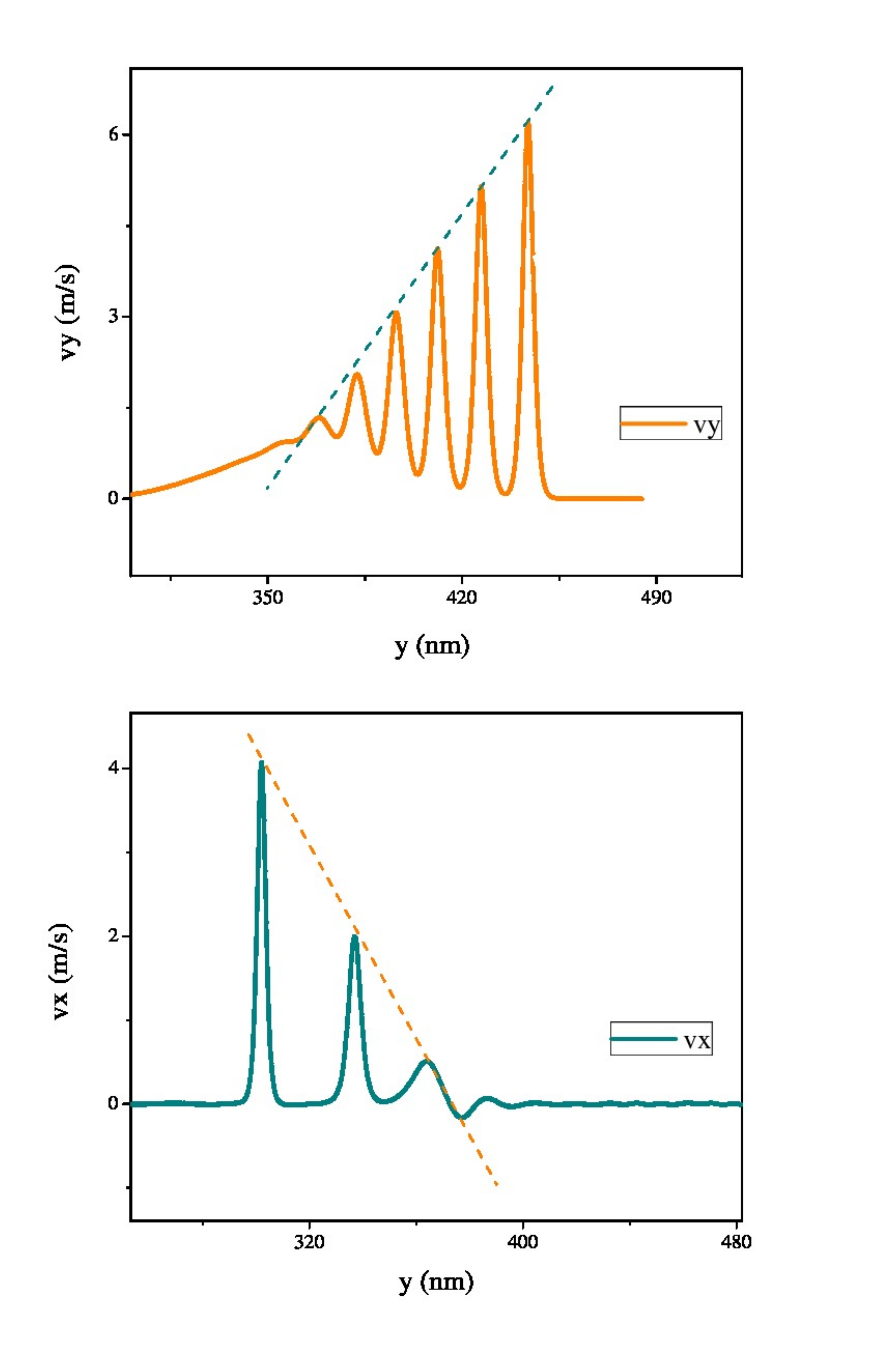}\caption{(Color online) Above it shows the longitudinal solitons are arranged
in a train with increasing amplitudes. Their amplitudes appear in
a straight line. Down it shows the transverse solitons are arranged
in a train with decreasing amplitudes. Their amplitudes also appear
in a straight line.}

\end{figure}

It is well known there are extreme difficulties to obtain a soliton
solution without making proper simplifications. Using different methods
and simplifications, different soliton solutions might be obtained.
To derive the soliton solution we use five simplifications. Thus it
is important to understand the conditions our simplifications correspond
to and the limits of these simplifications. 

The 1st simplification is to use the Brenner\textquoteright{}s potential
instead of the second-generation reactive empirical bond-order (REBO)
potential. As we point out both potentials describe the short-ranged
covalent-bonding C-C interactions of the carbon atoms in the same
way. Our simulation results are general for the potentials describing
the C-C interactions. Thus this simplification brings no significant
differences.

The 2nd simplification is to consider the displacements only in one
direction. It is due to the fact that only transverse/longitudinal
solitons would interact with each other while a transverse soliton
would not interact with a longitudinal soliton. It simplifies the
graphene flakes as a quasi-one dimensional chain. However in the molecular
dynamics simulations, displacements of all directions are considered.
Thus our analytical model is insufficient to describe the breather-like
solitons (Fig. 3) obtained in the simulations. In order to uncover
the origin of the breather-like soliton, a coupled NLS equation considering
both transverse and longitudinal displacements might be needed.

The 3rd simplification is to neglect the high-order harmonics in graphene.
It is due to the fact that the high-frequency modes are suppressed
in the molecular dynamics simulations by the cooling procedures. Here
it also explains why such strategy works. However high-order harmonics
may take part in the solitons obtained in the simulations as well.
Thus the single soliton solution might be just an approximation for
a multiple soliton solution. If it is so after a long time the solitons
may split into several secondary solitons if their velocities are
amplitude dependent. It brings us to the 4th simplification.

The 4th simplification is to neglect the velocity amplitude dependent
parameter. It is a valid simplification according to Fig. 7, Fig.
9 and Fig. 10. But should the velocities of the solitons be absolutely
amplitude independent or weakly amplitude dependent still remain unanswered.
To answer that question we need to investigate the long time behaviors
of the solitons. Thus we keep the solitons propagate as long as about
4 ns. The total propagation distance of the longitudinal solitons
is more than 0.1 mm. The total propagation distance of the transverse
solitons is more than 0.05 mm. It is shown in Fig. 18 the longtime
behavior of the longitudinal solitons in Graphene $\theta(0)$. It
is shown in Fig. 19 the longtime behavior of the transverse solitons
in Graphene $\theta(30)$. They gradually split into several secondary
solitons arranged in a train with increasing or decreasing amplitudes.
If the velocities are proportional to the amplitudes they would appear
with increasing amplitudes arrayed in a straight line. Meanwhile if
the velocities are inversely proportional to the amplitudes they would
appear with decreasing amplitudes arrayed in a straight line. So it
is shown in Fig. 20 the velocities of the longitudinal solitons are
proportional and the transverse solitons are inversely proportional
to the amplitudes.

The 5th simplification is to assume the soliton velocity is close
to the relative sound speed. The propagating velocity of the longitudinal
solitons is 20.0 km/s and the propagating velocity of the transverse
solitons is 11.4 km/s. Both velocities are close to the relative sound
speed which is 21.3 km/s and 13.6 km/s.

In summary, various solitons would emerge in the graphene flakes by
cooling procedures. They are subsonic transverse and longitudinal
solitons. They exhibit an averaged acceleration effect in collision
which means they are phase-shifted. A soliton solution is derived
by taking several simplifications which correspond to the simulation
conditions. The validity of the soliton solution is confirmed by comparing
with simulation results. When periodic boundary condition is used
in x-axis, the graphene flakes resemble simplified carbon nanotubes.
Similar solitons might be also important in carbon nanotubes. We hope
our work shed light on understanding the unusual thermal properties
of graphene and other similar materials and also benefit the development
of graphene based thermal device.
\begin{acknowledgments}
This work was supported by National Natural Science Foundation of
China (\#10925525 and \#10775115), and National Basic Research Program
of China (973 program) (\#2007CB814800).\end{acknowledgments}

\end{document}